\documentclass[10pt]{article}

\usepackage{amsfonts,amssymb,amsmath,mathrsfs,latexsym,bbm,dsfont,amsthm}
\usepackage{color}
\usepackage{graphicx}
\usepackage{pstricks,pstricks-add}
\usepackage{pst-plot}
\usepackage[scriptsize,nooneline,hang]{caption}
\usepackage[hang,nooneline,scriptsize]{subfigure}
\usepackage{rotating}
\usepackage[all]{xy}

\usepackage[abs]{overpic}

\begin{document}

\title{Geodesic Motion in the Singly Spinning Black Ring Spacetime}

\author{Saskia Grunau$^1$, Valeria Kagramanova$^1$, Jutta Kunz$^1$, Claus L\"ammerzahl$^{2,1}$\\
\vspace{0.1cm}\\
$^1$Institut f\"ur Physik, Universit\"at Oldenburg,
D--26111 Oldenburg, Germany \\
$^2$ZARM (Center of Applied Space Technology and Microgravity), Universit\"at Bremen, 
D--28359 Bremen, Germany
}


\maketitle

\begin{abstract}
We present analytical solutions of the geodesic equations of test particles and light in the five dimensional singly spinning black ring spacetime for special cases, since it does not appear possible to separate the Hamilton-Jacobi-equation for singly spinning black rings in general. Based on the study of the polynomials in the equations of motion we characterize the motion of test particles and light and discuss the associated orbits.
\end{abstract}

\section{Introduction}

Today String Theory is a promising candidate for the quantum theory of gravity. Since it requires more than four dimensions for its internal consistency, there has been growing interest in higher dimensional solutions and, in particular, in higher dimensional black holes (see e.g. \cite{Emparan:2008eg}).
The higher-dimensional generalizations of the stationary axisymmetric Kerr black holes  were found by Myers and Perry \cite{Myers:1986un},
who anticipated already the existence of higher-dimensional black rings.
In 2001 Emparan and Reall \cite{Emparan:2001wn} then found such black rings in five dimensions. They rotate in the direction of the ring and possess a horizon topology of $S^1 \times S^2$.
The phase diagram of these black rings together with the Myers-Perry black holes showed, that the uniqueness of 4-dimensional vacuum black holes does not generalize to higher dimensions.

In Myers-Perry black hole spacetimes the geodesic equations are separable \cite{Kubiznak:2006kt,Page:2006ka,Frolov:2006pe}.
However, the geodesic equations of five dimensional black rings do not seem to be separable, in general. Nevertheless, it is possible to separate the equations of motion on the rotational axis (which is actually a plane), in the equatorial plane and in the case $E=m=0$ (which is only possible in the ergosphere) \cite{Hoskisson:2007zk,Durkee:2008an}.

Hoskisson \cite{Hoskisson:2007zk} studied the geodesic motion of a singly spinning black ring and discussed the separability of the Hamilton-Jacobi equation, and provided numerical solutions. He analyzed numerically the motion on the rotational axis and the equatorial plane in detail. Some aspects of nullgeodesics in the equatorial plane were studied by Elvang, Emparan and Virmani \cite{Elvang:2006dd}.
The Pomerasky-Sen'kov doubly spinning black ring \cite{Pomeransky:2006bd} was studied  by Durkee \cite{Durkee:2008an}. He showed that it is possible to separate the Hamilton-Jacobi equation of the doubly spinning black ring in the case $E=m=0$ and analyzed the effective potential on the two axes (planes) of rotation and in the case $E=m=0$. The zero energy nullgeodesics of the singly spinning dipole black ring were analyzed by Armas \cite{Armas:2010pw}. In \cite{Igata:2010ye} Igata, Ishihara and Takamori concentrated on stable bound orbits in the singly spinning black ring spacetime, which they found numerically on and near the rotational axis.

So far the equations of motion for test particles in black ring spacetimes were only solved numerically, and no analytic solutions have been given.
The first to solve the geodesic equations in a black hole spacetime analytically was Hagihara \cite{Hagihara:1931}. He presented the solution of the geodesic equation for test particles in the Schwarzschild spacetime in terms of the elliptic Weierstra{\ss} $\wp$ function. 

When adding the cosmological constant to the Schwarzschild metric one encounters hyperelliptic curves in the geodesic equations. The equations of motion in Schwarzschild-(anti) de Sitter spacetimes in four dimensions were solved in \cite{Hackmann:2008zz}. Also analytical solutions of the geodesic equations in higher dimensional Myers-Perry spacetime \cite{Enolski:2010if} as well as in higher dimensional Schwarzschild, Schwarzschild-(anti) de Sitter, Reissner-Nordstr\"om and Reissner-Nordstr\"om-(anti) de Sitter \cite{Hackmann:2008tu} were found.

The mathematical method is based on the Jacobi inversion problem. The solution can be found if the problem is restricted to the Theta-divisor, the set of zeros of the theta function. Enolski, Pronine and Richter developed this method in 2003 to solve the problem of the double pendulum \cite{Enolski:2003}.

In this paper we present analytical solutions of the equations of motion of a singly spinning black ring. In the case $E=m=0$ and in the equatorial plane the equations are of elliptic type, however, on the rotational axis the geodesic equations are of hyperelliptic type.

\section{Singly Spinning Black Ring Spacetime}

\begin{figure}[h]
 \centering
 \includegraphics[width=12cm]{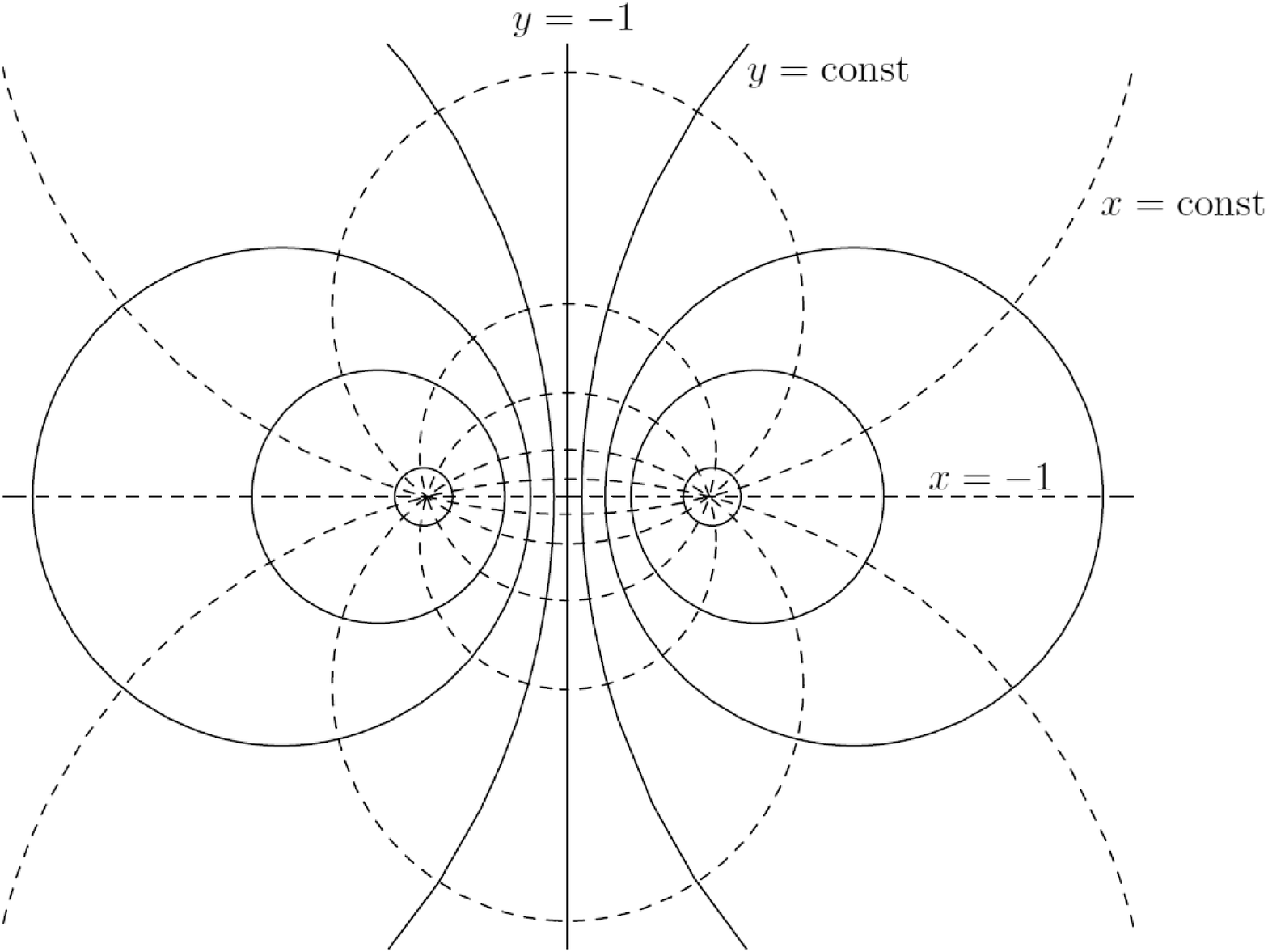}
 \caption{Toroidal coordinates (or ring coordinates) on a cross section at constant angles $\phi$ and $\psi$. Solid circles correspond to $y=\mathrm{const.}$ and dashed circles corredspond to  $x=\mathrm{const.}$. \cite{Emparan:2006mm}}
 \label{pic:ringcoord}
\end{figure}

The singly spinning black ring solution can be written in the form \cite{Durkee:2008an}
\begin{equation}
 \mathrm{d}s^2 = -\frac{H(y)}{H(x)}(\mathrm{d}t+\Omega_\psi \mathrm{d}\psi)^2 + \frac{R^2H(x)}{(x-y)^2} \left[ \frac{G(x)}{H(x)}\mathrm{d}\phi^2 + \frac{\mathrm{d}x^2}{G(x)} - \frac{G(y)}{H(y)}\mathrm{d}\psi^2 - \frac{\mathrm{d}y^2}{G(y)} \right] .
\end{equation}
The metric is given in toroidal coordinates (see \ref{pic:ringcoord}) where $-1 \leq x \leq 1$, $-\infty < y \leq -1$ and $-\infty < t < \infty$.  $\phi$ and $\psi$ are $2\pi$-periodic. The metric functions are
\begin{equation}
 \begin{split}
 G(x) &= (1-x^2)(1+\lambda x), \\
 H(x) &= 1+2x\lambda + \lambda ^2 , \\
 \Omega _\psi \mathrm{d}\psi &= -CR\frac{1+y}{H(y)} \mathrm{d}\psi , \quad \mathrm{where} \quad C^2\equiv 2\lambda ^2\frac{(1+\lambda)^3}{1-\lambda}.
 \end{split}
\end{equation}

The parameters $\lambda$ and $R$ describe the shape, mass and angular momentum of the ring. $\lambda$ lies in the range $0\leq \lambda <1$ to ensure the black ring is balanced.

A spacelike curvature singularity is located at $y=-\infty$. The metric has a coordinate singularity at $G(y)=0$, so the event horizon lies at $y_h=-\frac{1}{\lambda}$. The ergosphere of the singly spinning black ring is determined by $H(y)=0$, which is at $y_e=-\frac{1+\lambda^2}{2\lambda}$. Since we have $y_h<y_e<-1$, it is clear that an ergoregion does exist. The topology of the event horizon and the ergosphere is $S^1\times S^2$.\\

The inverse metric is
\begin{equation}
 \left(\frac{\partial}{\partial s}\right) ^2 = -\frac{H(x)}{H(y)} \left(\frac{\partial}{\partial t}\right) ^2 + \frac{(x-y)^2}{R^2H(x)}\left[ G(x) \left(\frac{\partial}{\partial x}\right) ^2 - G(y) \left(\frac{\partial}{\partial y}\right) ^2 + \frac{H(x)}{G(x)}\left(\frac{\partial}{\partial \phi}\right) ^2 - \frac{H(y)}{G(y)}\left( \frac{\partial}{\partial \psi} -\Omega_\psi \frac{\partial}{\partial t} \right)^2 \right] \, .
\end{equation}
The singly spinning black ring metric and its Hamiltonian $\mathscr{H} = \frac{1}{2}g^{ab}p_ap_b$ do not depend on the coordinates $t$, $\phi$ and $\psi$, so we have three conserved momenta $p_a=g_{ab}\dot{x}^b$ with the associated killing vector fields $\partial / \partial t$, $\partial / \partial \phi$ and $\partial / \partial \psi$. A dot denotes the derivative with respect to an affine parameter $\tau$.
\begin{eqnarray}
 -p_t &=& \frac{H(y)}{H(x)} (\dot{t}+\Omega_\psi\dot{\psi}) \equiv E \label{eqn:t-impuls}\\
 p_\phi &=& \frac{R^2G(x)}{(x-y)^2}\dot{\phi} \equiv \Phi \label{eqn:phi-impuls}\\
 p_\psi &=& -\Omega _\psi E  - \frac{R^2H(x)G(y)}{H(y)(x-y)^2}\dot{\psi} \equiv \Psi \, .\label{eqn:psi-impuls}
\end{eqnarray}
$E$ is the energy, $\Phi$ and $\Psi$ are the angular momenta in $\phi$- and $\psi$-direction. The conjugate momenta in $x$- and $y$-direction are:
\begin{eqnarray}
 p_x &=& \frac{R^2H(x)}{(x-y)^2G(x)}\dot{x} \label{eqn:x-impuls}\\
 p_y &=& -\frac{R^2H(x)}{(x-y)^2G(y)} \dot{y}\label{eqn:y-impuls}
\end{eqnarray}
To obtain the equations of motion for a particle in the singly spinning black ring spacetime we need the Hamilton-Jacobi equation:
\begin{equation}
 \frac{\partial S}{\partial \tau} + \mathscr{H} \left( x^a, \frac{\partial S}{\partial x^b}\right) = 0 \, .
 \label{eqn:hamilton-jacobi}
\end{equation}
We already have three constants of motion ($E$, $\Phi$ and $\Psi$) and the mass shell condition $g^{ab}p_a p_b=-m^2$ gives us a fourth , so we can make the ansatz
\begin{equation}
 S(\tau, t, x, y, \phi, \psi) = \frac{1}{2}m^2\tau -Et +\Phi\phi +\Psi\psi + S_x(x)+S_y(y) .
 \label{eqn:s-ansatz}
\end{equation}
Inserting this ansatz into (\ref{eqn:hamilton-jacobi}) gives
\begin{equation}
 0 = m^2 - \frac{H(x)}{H(y)}E^2 + \frac{(x-y)^2}{R^2H(x)}\left[ G(x) \left(\frac{\partial S}{\partial x}\right)^2 -G(y) \left(\frac{\partial S}{\partial y}\right)^2 + \frac{H(x)}{G(x)}\Phi^2 - \frac{H(y)}{G(y)}(\Psi + \Omega_\psi E)^2\right] \, .
\label{eqn:hjd-sring}
\end{equation}
The Hamilton-Jacobi equation does not seem to be separable in general. However, it is possible to separate the equation in the special case $E=m=0$. These zero energy null geodesics are only realisable in the ergoregion.

We can also obtain equations of motion for geodesics on the $\phi$- and $\psi$-axis by setting $x=\pm 1$ ($\phi$-axis) or $y=-1$ ($\psi$-axis). The plane  $x=\pm 1$ which is called the $\phi$-axis, is the equatorial plane of the black ring. The plane $y=-1$ which is called the $\psi$-axis corresponds to the rotational axis of the singly spinning black ring.\\

In the next sections we will study these special cases and solve the corresponding equations of motion analytically.

\section{Nullgeodesics in the Ergosphere}

For $E=m=0$ it is possible to separate the Hamilton-Jacobi equation:
\begin{equation}
 G(x) \left(\frac{\partial S}{\partial x}\right)^2 + \frac{H(x)}{G(x)}\Phi^2 = G(y) \left(\frac{\partial S}{\partial y}\right)^2 + \frac{H(y)}{G(y)}\Psi^2 \,.
\label{eqn:ham-jac-sing}
\end{equation}
With a separation constant $c$, the equation (\ref{eqn:ham-jac-sing}) splits into two:
\begin{eqnarray}
 G^2(x)\left(\frac{\partial S}{\partial x}\right)^2 &=& cG(x) -\Phi^2H(x) := X(x) \qquad \mathrm{and} \qquad \\
 G^2(y)\left(\frac{\partial S}{\partial y}\right)^2 &=& cG(y) -\Psi^2H(y) := Y(y),
\end{eqnarray}
so that
\begin{equation}
 S =\Phi\phi +\Psi\psi + \int \! \sqrt{X(x)} \, \mathrm{d}x + \int \! \sqrt{Y(y)} \, \mathrm{d}y .
\end{equation}
Using $p_a = \frac{\partial S}{\partial x^a}$ and (\ref{eqn:t-impuls})-(\ref{eqn:y-impuls}) the separated Hamilton-Jacobi equation gives the equations of motion:
\begin{eqnarray}
 \frac{\mathrm{d} x}{\mathrm{d} \gamma} &=& \sqrt{X(x)}  \label{eqn:sing-x-gleichung} \\
 \frac{\mathrm{d} y}{\mathrm{d} \gamma} &=& -\sqrt{Y(y)} \label{eqn:sing-y-gleichung} \\
 \frac{\mathrm{d} \phi}{\mathrm{d} \gamma} &=& \frac{H(x)\Phi}{G(x)} \label{eqn:sing-phi-gleichung} \\
 \frac{\mathrm{d} \psi}{\mathrm{d} \gamma} &=& -\frac{H(y)\Psi}{G(y)} \label{eqn:sing-psi-gleichung} \\
 \frac{\mathrm{d} t}{\mathrm{d} \gamma} &=& -\frac{CR(1+y)\Psi}{G(y)} \label{eqn:sing-t-gleichung}
\end{eqnarray}
where we have introduced the Mino-time \cite{Mino:2003yg} $\mathrm{d}\gamma= \frac{(x-y)^2}{R^2H(x)} \mathrm{d}\tau$.

\subsection{Classification of geodesics}

\begin{figure}
 \centering
 \includegraphics[width=8cm]{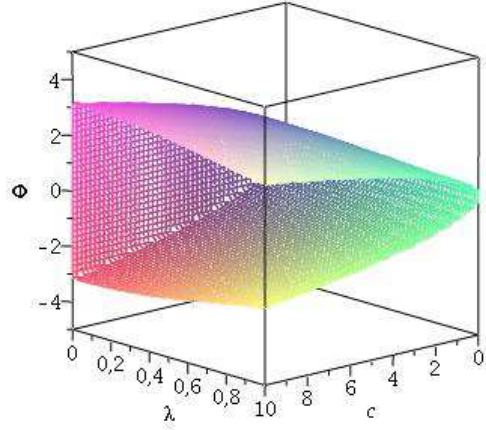}
 \caption{Three dimensional parametric $c$-$\lambda$-$\Phi$-diagram for the singly spinning black ring in the case $E=m=0$. Inside the structure $X(x)$ has two real zeros, outside the structure it has no real zeros in the allowed range of $x$.}
 \label{pic:sing-parameter}
\end{figure}

Equation (\ref{eqn:sing-x-gleichung}) and (\ref{eqn:sing-y-gleichung}) can be written as
\begin{eqnarray}
\left( \frac{\text{d}x}{\text{d}\gamma} \right) ^2 + U(x) &=& 0 \qquad \mathrm{where} \qquad  U(x)=\Phi^2H(x)-cG(x) \qquad \mathrm{and}, \\
\left( \frac{\text{d}y}{\text{d}\gamma} \right) ^2 + V(y) &=& 0 \qquad \mathrm{where} \qquad  V(y)=\Psi^2H(y)-cG(y).
\end{eqnarray}
$U(x)$ and $V(y)$ can be regarded as effective potentials (see \cite{Durkee:2008an}). To get real solutions for the $x$- and $y$-equation the effective potentials have to be negative. So $c\geq 0$ is required, because $H(x)\geq 0$ for $0\leq \lambda <1$ and $-1\leq x \leq 1$. The zeros of the effective potentials and hence $X$ and $Y$ mark the turning points of the motion of light or a test particle (in this case we only have light since $m=0$). A good way to determine the number of zeros are parametric diagrams. Figure \ref{pic:sing-parameter} shows a parametric $\Phi$-$\lambda$-$c$-diagram for $U(x)$. It turns out that $X(x)$ has two real zeros in the allowed range of $x$ or none. One zero is always negative while the other can be positive or negative. 

If $X(x)$ has two zeros the $x$-motion takes place between these two values, if $X(x)$ has a single zero at $x=0$ the $x$-motion stays constant. 

$Y(y)$ and accordingly $V(y)$ determine the type of the orbit. If $\lambda =0$ and $c\geq\Psi^2$, $Y(y)$ has no real zeros. Otherwise $Y(y)$ has always one real zero in the allowed range of $y$. That means the only possible orbit is a Terminating Orbit (TO), where light crosses the horizon and falls into the singularity.

\begin{figure}
 \centering
 \subfigure[Effective potential $U(x)$ for $\Phi =0.5$. There is one positive and one negative zero.]{
   \includegraphics[width=4.7cm]{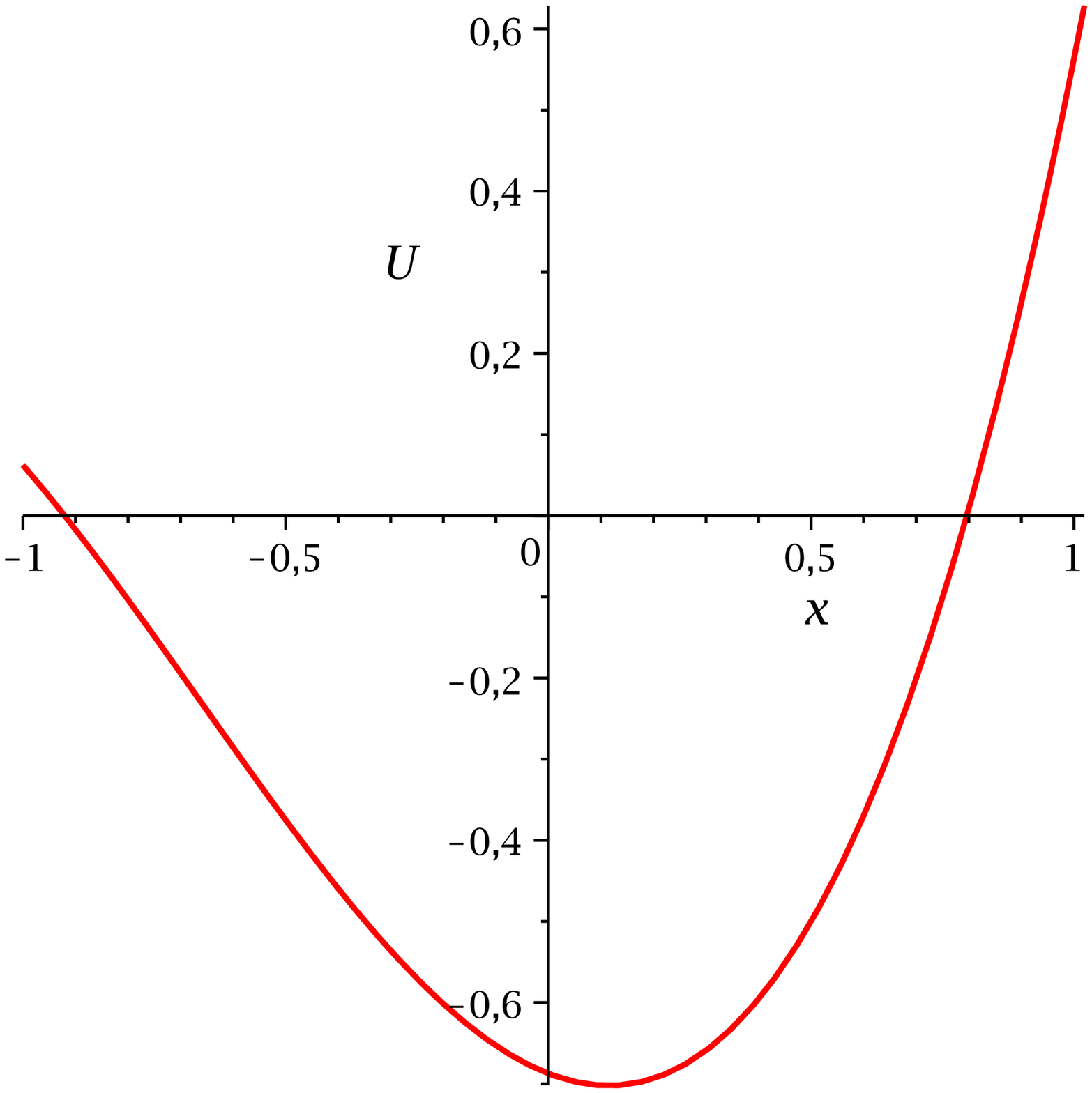}
  \label{pic:sing-potential1}
 }
 \subfigure[Effective potential $U(x)$ for $\Phi =0.9$. There are two negative zeros.]{
   \includegraphics[width=4.7cm]{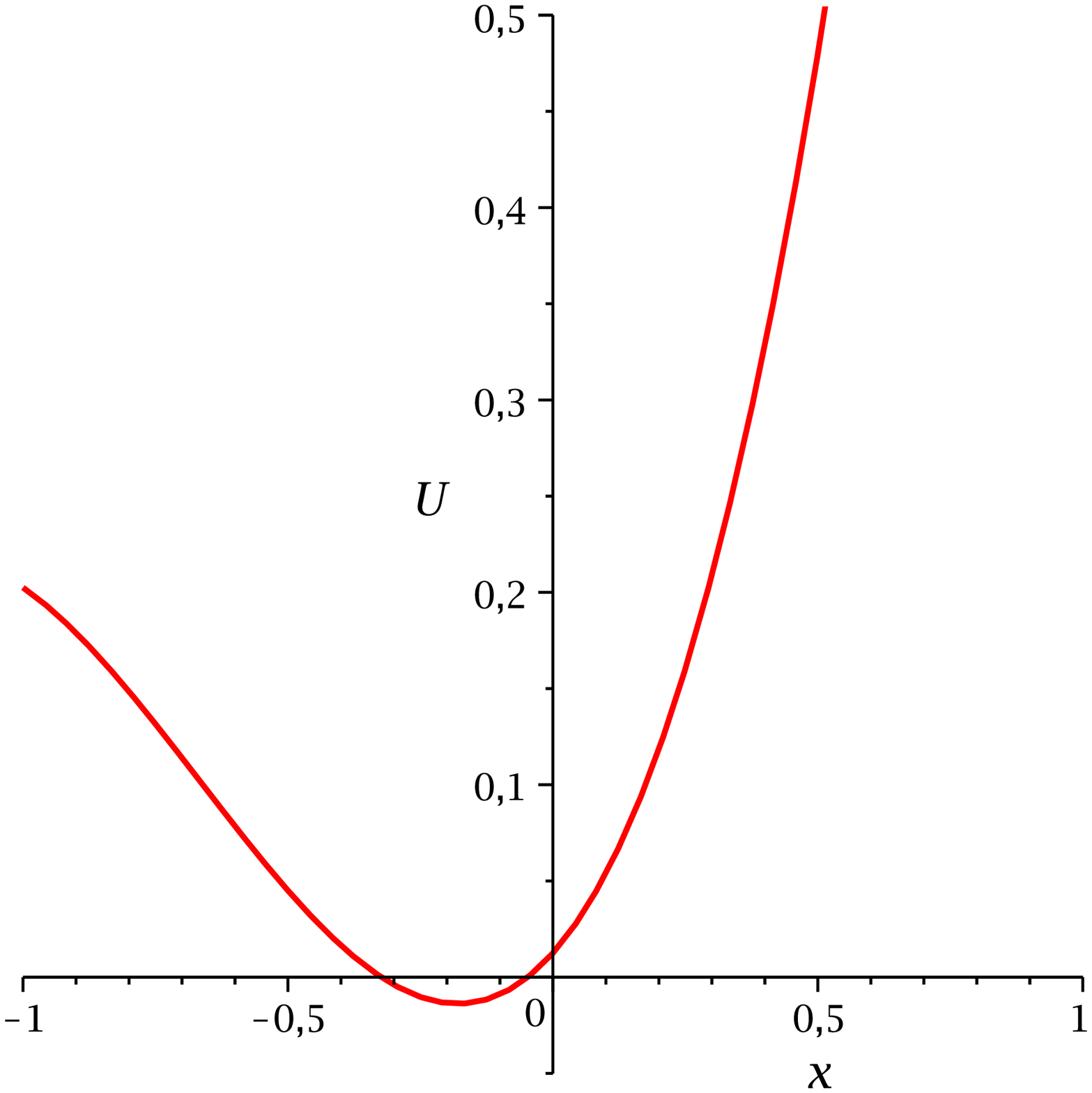}
  \label{pic:sing-potential2}
 }
 \subfigure[Effective potential $V(y)$ for $\Psi =5$. The point indicates the position of the turning point and the red horizontal dashed line shows the range of a terminating orbit. The horizon is marked by a vertical dashed line.]{
   \includegraphics[width=4.7cm]{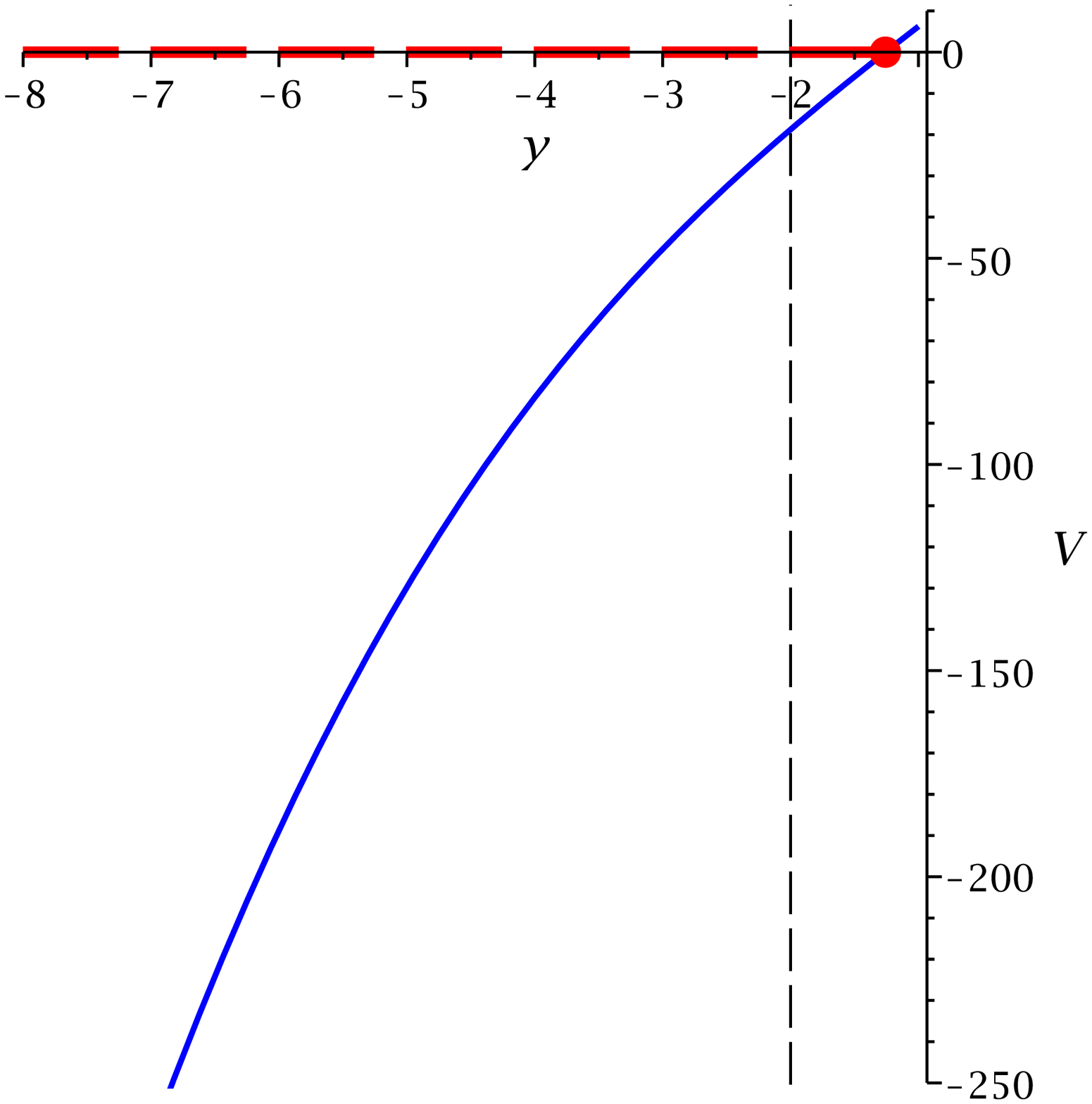}
 \label{pic:sing-potential3}
 }
 \caption{$c=1$,$\lambda=0.5$: Effective potentials for the singly spinning black ring in the case $E=m=0$.}
 \label{pic:sing-potential}
\end{figure}

See figure \ref{pic:sing-potential} for examples of the effective potentials.

\subsection{Solution of the $x$-equation}
\label{sec:ergo-xsol}

Equation (\ref{eqn:sing-x-gleichung}) can be written as 
\begin{equation}
 \left(\frac{\mathrm{d}x}{\mathrm{d}\gamma} \right) ^2 = X(x) = b_{x,3}x^3 +  b_{x,2}x^2 + b_{x,1}x + b_{x,0},
\end{equation}
where $X$ is a polynomial of third order with the coefficients
\begin{eqnarray}
 b_{x,3} &=& -c\lambda \nonumber\\
 b_{x,2} &=& -c \nonumber\\
 b_{x,1} &=& \lambda (c-2\Phi^2) \nonumber\\
 b_{x,0} &=& c-\Phi^2(1+\lambda^2) .
\end{eqnarray}
The substitution  $x=\frac{1}{b_{x,3}}\left( 4v-\frac{b_{x,2}}{3}\right) $ transforms the polynomial into the standard Weierstra{\ss} form
\begin{equation}
\left( \frac{\mathrm{d}v}{\mathrm{d}\gamma} \right) ^ 2= 4v^3 - g_{x,2} v -g_{x,3} :=P_{x,3}(v) ,
\label{eqn:sing-weierstrass-form}
\end{equation}
where 
\begin{equation}
g_{x,2}=\frac{b_{x,2}^2}{12} - \frac{b_{x,1} b_{x,3}}{4} \qquad \mathrm{and} \qquad
g_{x,3}=\frac{b_{x,1} b_{x,2} b_{x,3}}{48}-\frac{b_{x,0} b_{x,3}^2}{16}-\frac{b_{x,2}^3}{216} \ .
\end{equation}
Equation (\ref{eqn:sing-weierstrass-form}) is of elliptic type and is solved by the Weierstra{\ss} elliptic function \cite{Markushevich:1967}
\begin{equation}
v(\gamma )=\wp (\gamma - \gamma '_{\rm in},g_{x,2},g_{x,3}) \, , 
\end{equation}
where $\gamma' _{\rm in} = \gamma _{\rm in} + \int _{v_{x,\rm in}}^\infty \! \frac{\mathrm{d}v'}{\sqrt{4v'^3 - g_{x,2} v' -g_{x,3}}}$ and $v_{x,\rm in}=\frac{1}{4} \left( b_{x,3}x_{\rm in}+\frac{b_{x,2}}{3}\right) $.
Then the solution of (\ref{eqn:sing-x-gleichung}) takes the form
\begin{equation}
x (\gamma)=\frac{1}{b_{x,3}}\left[ 4\wp (\gamma - \gamma '_{\rm in},g_{x,2},g_{x,3}) -\frac{b_{x,2}}{3} \right] .
\end{equation}

\subsection{Solution of the $y$-equation}

Equation (\ref{eqn:sing-y-gleichung}) can be written as 
\begin{equation}
 \left(\frac{\mathrm{d}y}{\mathrm{d}\gamma} \right) ^2 =Y(y) = b_{y,3}y^3 +  b_{y,2}y^2 + b_{y,1}y + b_{y,0},
\end{equation}
where $Y$ is a polynomial of third order with the coefficients
\begin{eqnarray}
 b_{y,3} &=& -c\lambda \nonumber\\
 b_{y,2} &=& -c \nonumber\\
 b_{y,1} &=& \lambda (c-2\Psi^2) \nonumber\\
 b_{y,0} &=& c-\Psi^2(1+\lambda^2) .
\end{eqnarray}
The problem can be solved analogously to the $x$-equation. Here the solution is
\begin{equation}
y (\gamma)=\frac{1}{b_{y,3}}\left[ 4\wp (\gamma - \gamma ''_{\rm in},g_{y,2},g_{y,3}) -\frac{b_{y,2}}{3} \right] ,
\end{equation}
where  $\gamma'' _{\rm in} = \gamma _{\rm in} - \int _{v_{y,\rm in}}^\infty \! \frac{\mathrm{d}v'}{\sqrt{4v'^3 - g_{y,2} v' -g_{y,3}}} $ and $v_{y,\rm in}=\frac{1}{4} \left( b_{y,3}y_{\rm in}+\frac{b_{y,2}}{3}\right) $.

\subsection{Solution of the $\phi$-equation}
\label{sec:ergo-phisol}

Using (\ref{eqn:sing-x-gleichung}) the equation (\ref{eqn:sing-phi-gleichung}) becomes
\begin{equation}
 \mathrm{d}\phi = \frac{H(x)\Phi}{G(x)}\frac{\mathrm{d}x}{\sqrt{X(x)}} \qquad \mathrm{or}
\end{equation}
\begin{equation}
 \phi - \phi_{\rm in} = \Phi \int_{x_{\rm in}}^x \! \frac{H(x')\Phi}{G(x')} \, \frac{\mathrm{d}x'}{\sqrt{X(x')}} \, .
\end{equation}
We substitute $x=\frac{1}{b_{x,3}}\left( 4u -\frac{b_{x,2}}{3}\right)$ to transform $X(x)$ into the Weierstra{\ss} form $P_{x,3}$ (see (\ref{eqn:sing-weierstrass-form})):
\begin{equation}
 \phi - \phi_{\rm in} = \Phi \int_{u_{\rm in}}^u \! \frac{H\left( \frac{1}{b_{x,3}}\left( 4u' -\frac{b_{x,2}}{3}\right)\right) \Phi}{G\left( (\frac{1}{b_{x,3}}\left( 4u' -\frac{b_{x,2}}{3}\right)\right) } \, \frac{\mathrm{d}u'}{\sqrt{P_{x,3}(u')}}
\label{eqn:sing-Ix}
\end{equation}
Now $G$ has the zeros $p_{1,2}=\pm \frac{b_{x,3}}{4}+\frac{b_{x,2}}{12}$ and $p_3=-\frac{b_{x,3}}{4\lambda}+\frac{b_{x,2}}{12}$.
We next apply a partial fractions decomposition upon equation (\ref{eqn:sing-Ix}):
\begin{equation}
\phi - \phi_{\rm in} = \Phi \int^u_{u_{\rm in}} \sum^3_{j=1}\frac{H_j}{u-p_j} \frac{du'}{\sqrt{P_{x,3}(u')}}
\label{eqn:sing-Ix-partial}
\end{equation}
$H_j$ are constants which arise from the partial fractions decomposition and depend on the parameters of the metric and the test particle. Then we substitute $u = \wp (v, g_{x,2}, g_{x,3})$ with $\wp^\prime(v)=\sqrt{4 \wp^3(v)-g_{x,2}\wp(v)-g_{x,3}}$. Equation (\ref{eqn:sing-Ix-partial}) now simplifies to 
\begin{equation}
\phi - \phi_{\rm in} = \Phi \int^v_{v_{\rm in}} \sum^3_{j=1}\frac{H_j}{\wp(v)-p_j} dv
\end{equation}
with $v=v(\gamma)=\gamma-\gamma^\prime_{\rm in}$ and $v_{\rm in}=v(\gamma_{\rm in})$.

After solving the integrals of the third kind (see e.g. \cite{Enolski:2011id}), the final solution reads
\begin{equation}
 \phi (\gamma) = \Phi  \sum^3_{j=1} \frac{H_j}{\wp^\prime_x(v_{j})}\Biggl( 2\zeta_x(v_{j})(v-v_{\rm in}) + \log\frac{\sigma_x(v-v_{j})}{\sigma_x(v_{\rm in}-v_{j})} - \log\frac{\sigma_x(v+v_{j})}{\sigma_x(v_{\rm in}+v_{j})} \Biggr)  + \phi _{\rm in}
\end{equation}
with $p_j=\wp(v_j)$. The index $x$ refers to the Weierstra{\ss}-functions with respect to the parameters $g_{x,2}$ and $g_{x,3}$.

\subsection{Solution of the $\psi$-equation}
\label{sec:ergo-psisol}

Using (\ref{eqn:sing-y-gleichung}) equation (\ref{eqn:sing-psi-gleichung}) becomes
\begin{equation}
 \mathrm{d}\psi = \frac{H(y)\Psi}{G(y)}\frac{\mathrm{d}y}{\sqrt{Y(y)}} \qquad \mathrm{or}
\end{equation}
\begin{equation}
 \psi - \psi_{\rm in} = \Psi \int_{y_{\rm in}}^y \! \frac{H(y')\Phi}{G(y')} \, \frac{\mathrm{d}y'}{\sqrt{Y(y')}} \, .
\end{equation}
The $\psi$-equation can be solved analogously to the $\phi$-equation. With $v=v(\gamma)=\gamma-\gamma''_{\rm in}$, $v_{\rm in}=v(\gamma_{\rm in})$ and $p_j=\wp(v_j)$ the solution is
\begin{equation}
 \psi (\gamma) = \Psi  \sum^3_{j=1} \frac{K_j}{\wp^\prime_y(v_{j})}\Biggl( 2\zeta_y(v_{j})(v-v_{\rm in}) + \log\frac{\sigma_y(v-v_{j})}{\sigma_y(v_{\rm in}-v_{j})} - \log\frac{\sigma_y(v+v_{j})}{\sigma_y(v_{\rm in}+v_{j})} \Biggr) + \psi _{\rm in} \, .
\end{equation}
$K_j$ are constants which arise from the partial fractions decomposition and depend on the parameters of the metric and the test particle. The index $y$ refers to the Weierstra{\ss}-functions with respect to the parameters $g_{y,2}$ and $g_{y,3}$.

\subsection{Solution of the $t$-equation}

Using (\ref{eqn:sing-y-gleichung}) we can write (\ref{eqn:sing-t-gleichung}) as
\begin{equation}
 \mathrm{d}t = CR\Psi\frac{1+y}{G(y)} \frac{\mathrm{d}y}{\sqrt{Y(y)}} = CR\Psi\frac{1}{(1-y)(1+\lambda y)} \frac{\mathrm{d}y}{\sqrt{Y(y)}} \qquad \mathrm{or}
\end{equation}
\begin{equation}
 t - t_{\rm in} = CR\Psi \int_{y_{\rm in}}^y \! \frac{1}{(1-y')(1+\lambda y')} \, \frac{\mathrm{d}y'}{\sqrt{Y(y')}} \, .
\label{eqn:sing-em0-tint}
\end{equation}
The $t$-equation can be solved in an analogous way to the $\phi$- and $\psi$-equation. We substitute $y=\frac{1}{b_{y,3}}\left( 4u -\frac{b_{y,2}}{3}\right)$. The integral (\ref{eqn:sing-em0-tint}) is of the third kind and has the poles  $p_1=\frac{b_{x,3}}{4}+\frac{b_{x,2}}{12}$ and $p_2=-\frac{b_{x,3}}{4\lambda}+\frac{b_{x,2}}{12}$ (with respect to $u$). Then we apply a partial fractions decomposition, where the constants $M_j$ arise and we substitute again  $u = \wp (v, g_{y,2}, g_{y,3})$. After the solution of the occurring elliptic integrals of the third kind, the solution of (\ref{eqn:sing-t-gleichung}) yields
\begin{equation}
 t (\gamma) = CR\Psi  \sum^2_{j=1} \frac{M_j}{\wp^\prime_y(v_{j})}\Biggl( 2\zeta_y(v_{j})(v-v_{\rm in}) + \log\frac{\sigma_y(v-v_{j})}{\sigma_y(v_{\rm in}-v_{j})} - \log\frac{\sigma_y(v+v_{j})}{\sigma_y(v_{\rm in}+v_{j})} \Biggr)  + t _{\rm in} \, .
\end{equation}

\subsection{The Orbits}
\label{sec:em0-orbit}

\begin{figure}[h!]
 \centering
 \includegraphics[width=10cm]{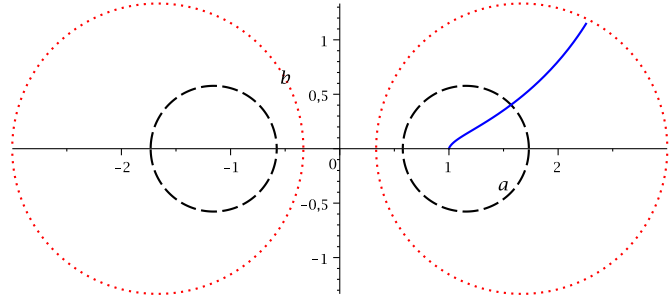}
 \caption{$c=1$, $\lambda=0.5$, $\Phi=0.5$ and $\Psi=5$\newline
  $a$-$b$-plot of a TO for the singly spinning black ring in the case $E=m=0$. The black dashed circles are the event horizon and the red dotted circles denote the ergosphere.}
 \label{pic:sing-to}
\end{figure}

In the ergosphere of a singly spinning black ring only TOs with $E=m=0$ are possible. Figure \ref{pic:sing-to} shows a TO plotted in the $x$-$y$-plane in Cartesian coordinates ($a$, $b$).

To change from the ring coordinates to the polar coordinates ($\rho$, $\theta$) the transformation
\begin{equation}
 \rho=\frac{R\sqrt{y^2-x^2}}{x-y}\, ,\quad \tan \theta =\sqrt{\frac{y^2-1}{1-x^2}}
\end{equation}
is used. Then conventional Cartesian coordinates take the form
\begin{equation}
 a = \rho\sin\theta \, ,\quad b = \rho\cos\theta
\end{equation}
(see \cite{Hoskisson:2007zk} or \cite{Lim:2008}). The singularity of the black ring is at $a=\pm 1$, $b=0$.

\section{Geodesics on the rotational axis}

The surface $y=-1$ is the axis of  rotation of the singly spinning black ring. Here the Hamilton-Jacobi equation depends on the coordinate $x$ only. We set $y=-1$, $\Psi =0$ and $p_y=\frac{\partial S}{\partial y}=0$ in the Hamilton-Jacobi equation (\ref{eqn:hjd-sring}):
\begin{equation}
 0 = m^2 - \frac{H(x)}{(1-\lambda)^2}E^2+\frac{(x+1)^2}{R^2H(x)}\left[ G(x) \left(\frac{\partial S}{\partial x}\right) ^2 + \frac{H(x)}{G(x)}\Phi^2\right] \, .
\end{equation}
This can be rearranged to
\begin{equation}
 \left(\frac{\partial S}{\partial x}\right) ^2 = \frac{R^2H(x)}{(x+1)^2G(x)}\left[\frac{H(x)}{(1-\lambda)^2}E^2-m^2 \right] - \frac{H(x)}{G^2(x)}\Phi^2 := X_S .
\end{equation}
Then we have
\begin{equation}
 S=\frac{1}{2}m^2\tau -Et+\Phi\phi + \int\! \sqrt{X_S} \, \mathrm{d}x \, .
\end{equation}
Now we set the partial derivatives of $S$ with respect to the constants $m^2$, $E$ and $\Phi$ to zero in order to obtain the equations of motion.
With the Mino-time \cite{Mino:2003yg} $\mathrm{d}\gamma=\frac{x+1}{R^2H(x)}\mathrm{d}\tau$ the equations of motion take the form
\begin{eqnarray}
 \frac{\mathrm{d}x}{\mathrm{d}\gamma} &=& \left\lbrace R^2H(x)G(x)\left( H(x)\frac{E^2}{(1-\lambda)^2}-m^2 \right) -\Phi^2 H(x) (x+1)^2 \right\rbrace ^{1/2} \nonumber\\
      &:=& \sqrt{X(x)} \label{eqn:sing-psi-x-gleichung} \, ,\\
 \frac{\mathrm{d}\phi}{\mathrm{d}\gamma} &=& \Phi\frac{(x+1) H(x)}{G(x)} \label{eqn:sing-psi-phi-gleichung} \, ,\\
 \frac{\mathrm{d}t}{\mathrm{d}\gamma} &=&\frac{R^2EH^2(x)}{(x+1)(1-\lambda)^2} \label{eqn:sing-psi-t-gleichung} \, ,
\end{eqnarray}
where $X(x)$ is a polynomial of fifth order.

\subsection{Classification of geodesics}

From (\ref{eqn:sing-psi-x-gleichung}) we can read off the effective potential consisting of two parts $U_+(x)$ and $U_-(x)$ (to be consistent with the effective potential on the equatorial plane later on):
\begin{equation}
 X=a(x)(E-U_+)(E-U_-) \, .
\end{equation}
Since $X(x)$ can be written as $X(x)=a(x)E^2+b(x)$ the effective potential takes the form
\begin{equation}
 U_\pm (x) = \pm \sqrt{-\frac{b(x)}{a(x)}},
\end{equation}
where $a(x)=\frac{R^2H^2(x)G(x)}{(1-\lambda)^2}$ and $b(x)=-R^2H(x)G(x)m^2-\Phi^2H(x)(x+1)^2$.

\begin{figure}[h]
 \centering
 \subfigure[$R=1$, $m=1$, $\lambda=0.4$ and $\Phi=0$ \newline Examples of orbits of type A and B. There are none or one turning points.]{
   \includegraphics[width=4.5cm]{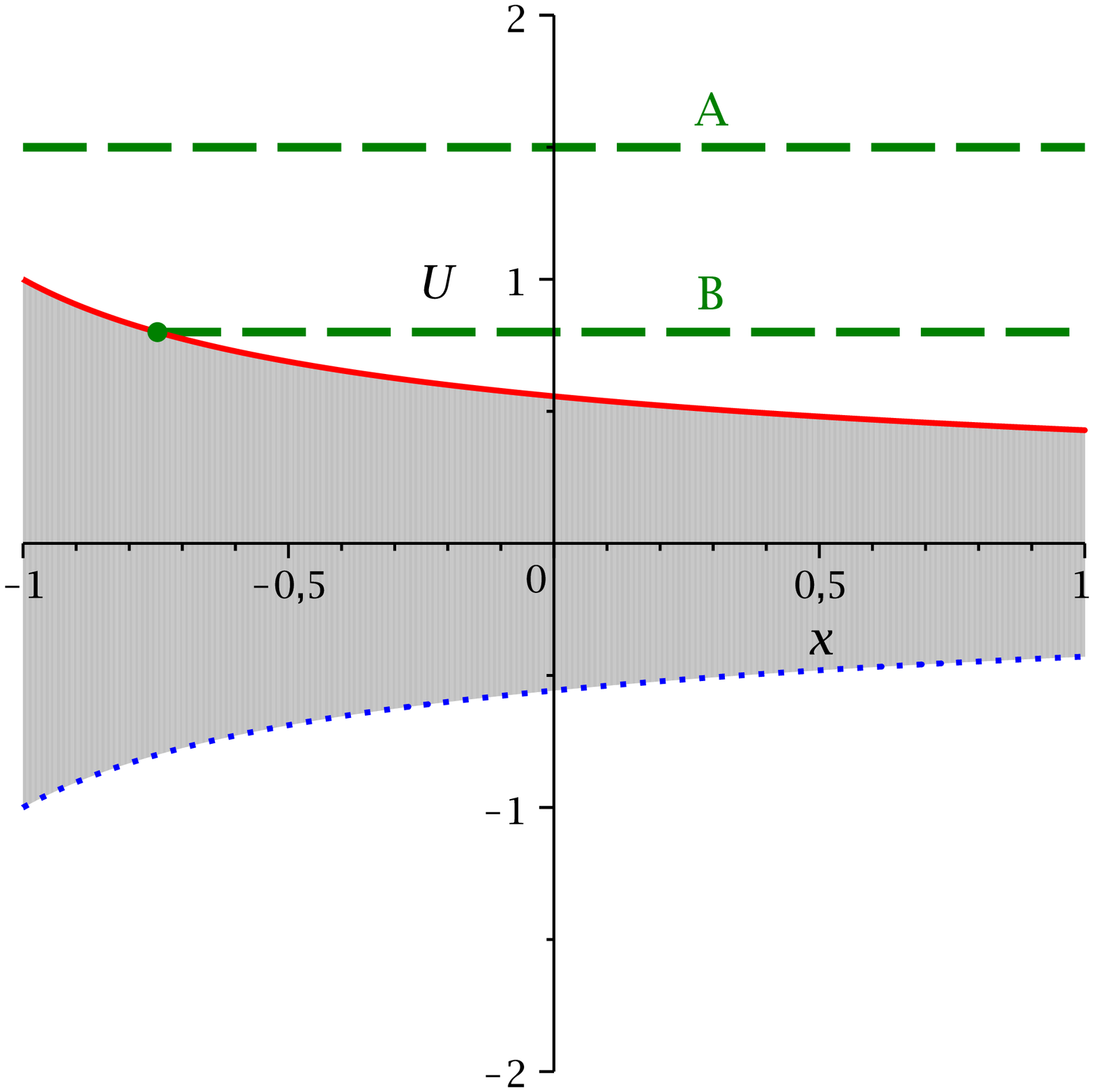}
 }
 \subfigure[$R=1$, $m=1$, $\lambda=0.4$ and $\Phi=1$ \newline Examples of orbits of type C and D. If $|\Phi|>0$ there is a potential barrier which prevents the geodesics from reaching $x=+1$. There are one or two turning points.]{
   \includegraphics[width=4.5cm]{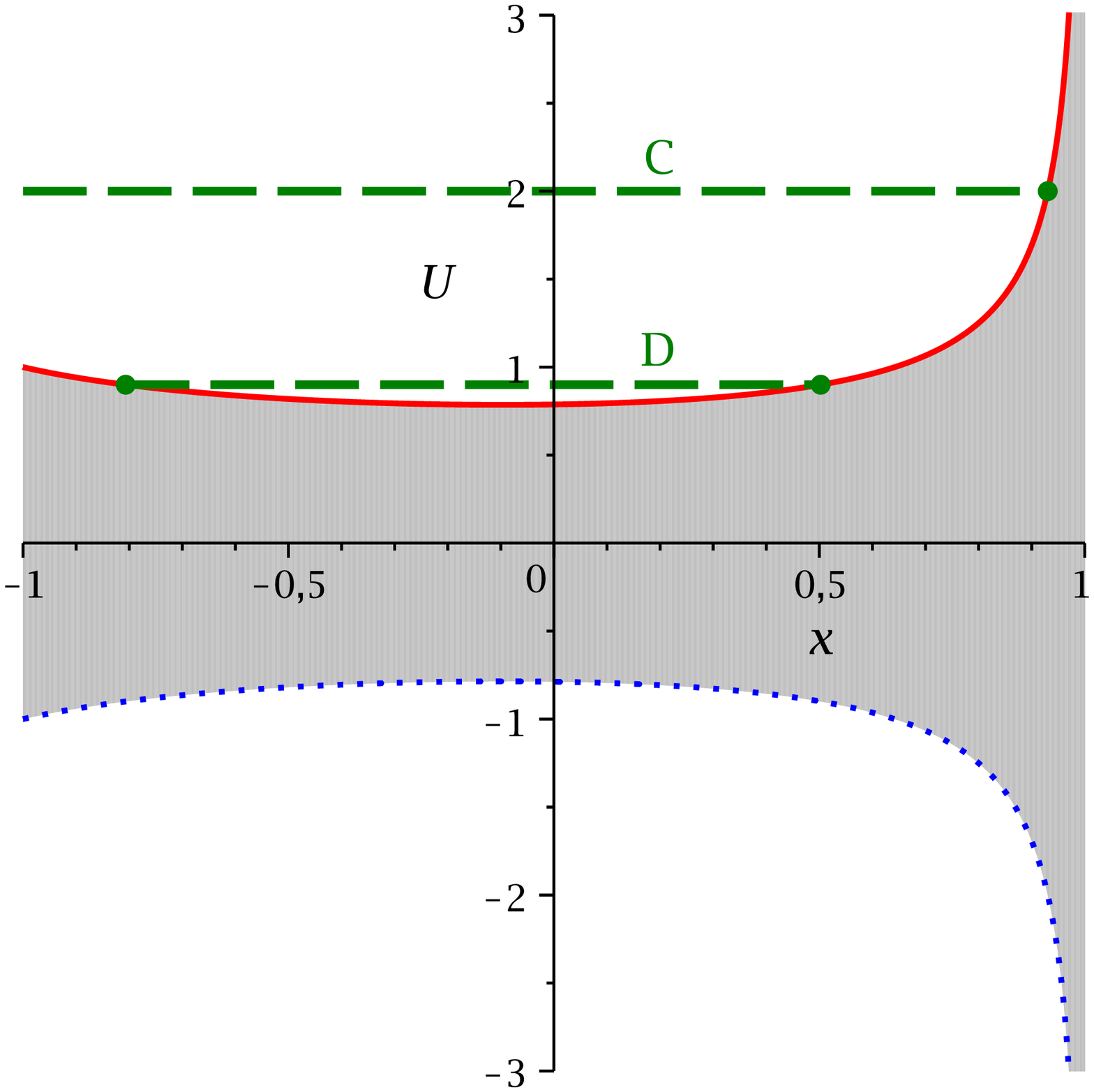}
 }
 \subfigure[$R=1$, $m=1$, $\lambda=0.8$ and $\Phi=8$ \newline Example of an orbit of type E. The potential can have lokal extrema which lead to three turning points.]{
   \includegraphics[width=4.5cm]{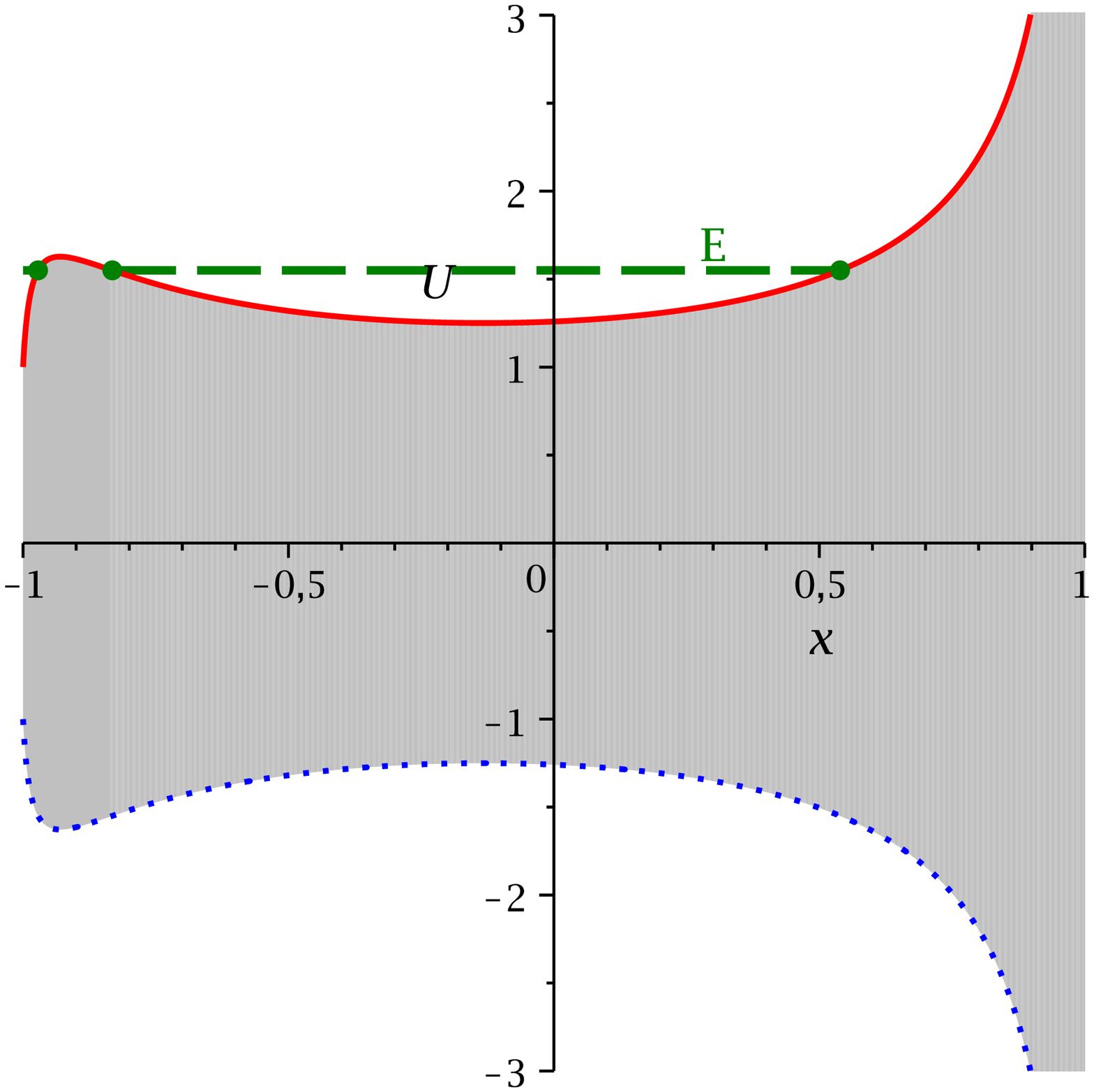}
 }
 \caption{Effective potentials $U_+(x)$ (red, solid) and $U_-(x)$ (blue, dotted) on the $\psi$ axis of the ring. The grey area is a forbidden zone, where no motion is possible. Green dashed lines represent energies and green points mark the turning points.}
 \label{pic:s-psi-orbits1}
\end{figure}

Figure \ref{pic:s-psi-orbits1} shows the effective potential for the motion on the $\psi$ axis. $U_+$ is plotted in red (solid) while $U_-$ is plotted in blue (dotted). The grey area between the two parts of the potential is a forbidden zone where no motion is possible because $X(x)$ becomes negative there. $U_+$ and $U_-$ are symmetric ($U_+=-U_-$) and meet at the horizon.

$x=-1$ is always a zero of $X(x)$, but since the point $x=-1$, $y=-1$ corresponds to infinity in cartesian coordinates it is not a real turning point of the test particle. If $\Phi=0$ then $X(x)$ has the zeros $x=-1$ and $x=+1$ and possibly a third zero between -1 and +1. The coordinate range of $x$ ($-1\leq x\leq +1$) only covers the space from infinity (-1) to the center of the black ring (+1). Since there is no potential barrier at $x=+1$ for $\Phi=0$, light and test particles with the right amount of energy cross the center of the black ring and continue their orbit at the other side of the black ring.

If $\Phi=0$ then none or one turning point exists. If $|\Phi|>0$ there is a potential barrier which prevents the geodesics from reaching $x=+1$ and one or two turning points exist. For larger $\lambda$ and $|\Phi|$ the potential has local extrema which lead to three turning points.

Possible orbits are Bound Orbits (BO), where light or test particles circle the black ring, and Escape Orbits (EO), where light or test particles approach the black ring, turn around at a certain point and escape the gravitational field.

There are five different types of orbits (see table \ref{tab:s-psi-typen-orbits1}).
\begin{itemize}
 \item Type A:\\
  $X(x)$ has no zero in the range  $-1<x<1$. EOs without a turning point exist. The orbit crosses the equatorial plane ($x=+1$) and reaches infinity ($x=-1$ and $y=-1$).
 \item Type B:\\
  $X(x)$ has one zero in the range  $-1<x<1$. BOs with a turning point on each side of the ring exist, so that the orbit crosses the equatorial plane ($x=+1$).
 \item Type C:\\
  $X(x)$ has one zero in the range  $-1<x<1$. EOs with a turning point exist.
 \item Type D:\\
  $X(x)$ has two zeros in the range  $-1<x<1$. BOs which do not cross the equatorial plane exist.
 \item Type E:\\
  $X(x)$ has three zeros in the range  $-1<x<1$. BOs which do not cross the equatorial plane and EOs exist.
\end{itemize}

\begin{table}[h]
\begin{center}
\begin{tabular}{|lcll|}\hline
type &  zeros  & range of $x$ & orbit \\
\hline\hline
A & 0 & 
\begin{pspicture}(-2.5,-0.2)(3,0.2)
\psline[linewidth=0.5pt]{->|}(-2.5,0)(3,0)
\psline[linewidth=1.2pt]{-}(-2.5,0)(3,0)
\end{pspicture}
  & EO
\\  \hline
B & 1 & 
\begin{pspicture}(-2.5,-0.2)(3,0.2)
\psline[linewidth=0.5pt]{->|}(-2.5,0)(3,0)
\psline[linewidth=1.2pt]{*-}(-1.5,0)(3,0)
\end{pspicture}
  & BO 
\\ \hline
C  & 1 &
\begin{pspicture}(-2.5,-0.2)(3,0.2)
\psline[linewidth=0.5pt]{->|}(-2.5,0)(3,0)
\psline[linewidth=1.2pt]{-*}(-2.5,0)(2,0)
\end{pspicture}
& EO 
\\ \hline
D & 2 & 
\begin{pspicture}(-2.5,-0.2)(3,0.2)
\psline[linewidth=0.5pt]{->|}(-2.5,0)(3,0)
\psline[linewidth=1.2pt]{*-*}(-1,0)(2,0)
\end{pspicture}
  & BO
\\ \hline
E & 3 & 
\begin{pspicture}(-2.5,-0.2)(3,0.2)
\psline[linewidth=0.5pt]{->|}(-2.5,0)(3,0)
\psline[linewidth=1.2pt]{*-*}(-0,0)(2,0)
\psline[linewidth=1.2pt]{-*}(-2.5,0)(-1.0,0)
\end{pspicture}
& EO, BO 
\\ \hline\hline
\end{tabular}
\caption{Types of orbits of light and particles in the singly spinning black ring spacetime for $y=-1$, $\Psi =0$. The thick lines represent the range of the orbits. The turning points are shown by thick dots. The number of zeros in the table is the number of zeros in the range $-1<x<1$. }
\label{tab:s-psi-typen-orbits1}
\end{center}
\end{table}

\subsection{Solution of the $x$-equation}
\label{sec:sing-x-solution}

Equation (\ref{eqn:sing-psi-x-gleichung}) can be written as
\begin{equation}
 \left( \frac{\mathrm{d}x}{\mathrm{d}\gamma}\right)^2 = X(x) = a_{x,5} x^5 + a_{x,4} x^4 + a_{x,3} x^3 + a_{x,2} x^2 + a_{x,1} x + a_{x,0},
 \label{eqn:sing-psi-x-quadrgleichung}
\end{equation}
where $X(x)$ is a polynomial of fifth order with the coefficients
\begin{eqnarray}
 a_{x,5} &=& \frac{-4 R^2 \lambda^3 E^2}{(1-\lambda)^2} \nonumber\\
 a_{x,4} &=& \frac{[-2R^2\lambda-R^2 (1+\lambda^2)\lambda]2\lambda E^2}{(1-\lambda)^2 } - 2R^2\lambda^2 \left( \frac{(1+\lambda^2)}{(1-\lambda)^2}E^2 -m^2 \right) \nonumber\\
 a_{x,3} &=& 2\Phi^2\lambda + \frac{[2R^2\lambda^2-R^2(1+\lambda^2)]2\lambda E^2}{(1-\lambda)^2} + (-2R^2\lambda-R^2(1+\lambda^2)\lambda) \left( \frac{(1+\lambda^2)}{(1-\lambda)^2}E^2 -m^2 \right) \nonumber\\
 a_{x,2} &=& -4 \Phi^2 \lambda-\Phi^2 (1+\lambda^2) + \frac{[2R^2\lambda^2+R^2(1+\lambda^2)]2\lambda E^2}{(1-\lambda)^2} + (2R^2\lambda-R^2(1+\lambda^2)) \left( \frac{(1+\lambda^2)}{(1-\lambda)^2}E^2 -m^2 \right) \nonumber\\
 a_{x,1} &=& -2\Phi^2\lambda-2\Phi^2(1+\lambda^2) + \frac{2R^2(1+\lambda^2)\lambda E^2}{(1-\lambda)^2} + (2R^2\lambda + R^2(1+\lambda^2)\lambda  ) \left( \frac{(1+\lambda^2)}{(1-\lambda)^2}E^2 -m^2 \right) \nonumber\\
 a_{x,0} &=& -\Phi^2(1+\lambda^2) + R^2(1+\lambda^2)\left( \frac{(1+\lambda^2)}{(1-\lambda)^2}E^2 -m^2 \right) .
\label{eqn:sing-xkoeff}
\end{eqnarray}
A separation of variables gives the hyperelliptic integral
\begin{equation}
 \gamma - \gamma_{\rm in} = \int _{x_{\rm in}}^x \! \frac{\mathrm{d}x'}{\sqrt{X(x')}}.
 \label{eqn:sing-psi-x-intgleichung}
\end{equation}
A canonical basis of holomorphic ($\mathrm{d}u_i$) and meromorphic ($\mathrm{d}r_i$) differentials associated with the hyperelliptic curve $w^2=X(x)$ is given by (see \cite{Enolski:2011id} or \cite{Hackmann:2008tu})
\begin{equation}
 \mathrm{d}u_1 := \frac{\mathrm{d}x}{\sqrt{X(x)}}, \qquad \mathrm{d}u_2 := \frac{x\mathrm{d}x}{\sqrt{X(x)}},
 \label{eqn:sing-dz}
\end{equation}
\begin{equation}
 \mathrm{d}r_1 := (3a_{x,5}x^3+2a_{x,4}x^2+a_{x,3}x)\frac{\mathrm{d}x}{4\sqrt{X(x)}}, \qquad \mathrm{d}r_2 := a_{x,5}x^2\frac{\mathrm{d}x}{4\sqrt{X(x)}} \, .
 \label{eqn:sing-dr}
\end{equation}
Furthermore we introduce the holomorphic and meromorphic period matrices $(2\omega, 2\omega ')$ and $(2\eta, 2\eta ')$:
\begin{equation}
 \begin{split}
  2\omega_{ij} := \oint_{a_j} \mathrm{d}u_i, \qquad 2\omega'_{ij} := \oint_{b_j} \mathrm{d}u_i, \\
  2\eta_{ij} := -\oint_{a_j} \mathrm{d}r_i, \qquad 2\eta'_{ij} := -\oint_{b_j} \mathrm{d}r_i ,
 \end{split}
\end{equation}
with $i,j = 1,2$, where $\{ a_1,a_2;b_1,b_2 \}$ is the canonical basis of closed paths.
The normalized holomorphic differentials are
\begin{equation}
 \mathrm{d}\boldsymbol{v} := (2\omega)^{-1}\mathrm{d}\boldsymbol{u}, \qquad \mathrm{d}\boldsymbol{u}=
 \left(
 \begin{array}{c}
  \mathrm{d}u_1\\
  \mathrm{d}u_2\\
 \end{array}
 \right) .
\end{equation}
The solution of equation (\ref{eqn:sing-psi-x-quadrgleichung}) is extensively discussed in \cite{Hackmann:2008zz,Enolski:2010if,Hackmann:2008tu,Enolski:2011id}, and is given by the derivatives $\sigma_i$ of the Kleinian sigma function $\sigma (\boldsymbol{u}) = k e^{-(1/2)\boldsymbol{u}^t\eta\omega^{-1}\boldsymbol{u}} \vartheta ((2\omega)^{-1}\boldsymbol{u} + \boldsymbol{K}_{x_{\rm in}};\tau)$: 
\begin{equation}
 x(\gamma)=-\frac{\sigma_1(\boldsymbol{\gamma}_\Theta)}{\sigma_2(\boldsymbol{\gamma}_\Theta)} \, ,
\end{equation}
where
\begin{equation}
 \boldsymbol{\gamma}_\Theta :=
 \left(
 \begin{array}{c}
 \gamma - \gamma_{\rm in}' \\
 \gamma_2
 \end{array}
 \right).
\end{equation}
The constant $\gamma_{\rm in}' = \gamma_{\rm in} +\int_{x_{\rm in}}^{\infty}\! \mathrm{d}u_1$ depends on $ \gamma_{\rm in}$ and $x_{\rm in}$ only. $\gamma_2$ is defined by the vanishing condition of the Kleinian sigma function $\sigma (\boldsymbol{\gamma}_\Theta) = 0$ so that $(2\omega )^{-1}\boldsymbol{\gamma}_\Theta$ is an element of the theta divisor $\Theta_{ \boldsymbol{K_\infty}}$ (the set of zeros of the theta function) where
\begin{equation}
 \boldsymbol{K}_\infty = \tau 
 \left(
 \begin{array}{c}
  1/2\\
  1/2\\
 \end{array}
 \right) +
 \left(
 \begin{array}{c}
  0\\
  1/2\\
 \end{array}
 \right)
\end{equation}
is the vector of Riemann constants and $\tau$ is the Riemann period matrix defined as $\tau := \omega ^{-1}\omega'$.

\subsection{Solution of the $\phi$-equation}
\label{sec:sing-psi-phi-lösung}

With (\ref{eqn:sing-psi-x-gleichung}) equation (\ref{eqn:sing-psi-phi-gleichung}) yields
\begin{equation}
\mathrm{d}\phi = \Phi \frac{(x+1)H(x)}{G(x)} \frac{\mathrm{d}x}{\sqrt{X(x)}} = \Phi \frac{H(x)}{(1-x)(1+\lambda x)} \frac{\mathrm{d}x}{\sqrt{X(x)}}
\end{equation}
or
\begin{equation}
 \phi - \phi_{\rm in} = \Phi \int _{x_{\rm in}}^x \! \frac{H(x')}{(1-x')(1+\lambda x')} \, \frac{\mathrm{d}x'}{\sqrt{X(x')}} \, .
\label{eqn:psi-phiint}
\end{equation}
The integral (\ref{eqn:psi-phiint}) has poles at $p_1=1$ and $p_2=-\frac{1}{\lambda}$.

Now we apply a partial fractions decomposition upon (\ref{eqn:psi-phiint}):
\begin{equation}
 \phi - \phi_{\rm in} = \Phi \int _{x_{\rm in}}^x \! \sum _{i=1}^2 \frac{K_i}{x'-p_i}  \, \frac{\mathrm{d}x'}{\sqrt{X(x')}} \, ,
\end{equation}
where $K_i=\mp\lambda -1$ are constants which arise from the partial fractions decomposition.

The differentials in the equation above are of the third kind and can be solved with the help of the following equation (see \cite{Enolski:2011id}). 
\begin{align}\begin{split}
W\int_{P'}^P\frac{1}{x-Z}\frac{\mathrm{d}x}{w} = & 2\int_{P'}^P \mathrm{d}\boldsymbol{u}^T(x,y) \left[ \boldsymbol{\zeta} \left( \int_{(e_2,0)}^{(Z,W)} \mathrm{d} \boldsymbol{u} + \boldsymbol{K}_\infty  \right) - 2( \boldsymbol{\eta}^{\prime}\boldsymbol{\varepsilon}^\prime + \boldsymbol{\eta}\boldsymbol{\varepsilon} )  - \frac12 \boldsymbol{\mathfrak{Z}}(Z,W)    \right]\\
& +\ln \frac{\sigma\left(\int_{\infty}^P \mathrm{d}\boldsymbol{u}- \int_{(e_2,0)}^{(Z,W)} \mathrm{d}\boldsymbol{u} - \boldsymbol{K}_\infty  \right)}{\sigma\left(\int_{\infty}^P \mathrm{d}\boldsymbol{u}+ \int_{(e_2,0)}^{(Z,W)} \mathrm{d}\boldsymbol{u} - \boldsymbol{K}_\infty \right)}
- \mathrm{ln} \frac{\sigma\left(\int_{\infty}^{P'} \mathrm{d}\boldsymbol{u} - \int_{(e_2,0)}^{(Z,W)} \mathrm{d}\boldsymbol{u} - \boldsymbol{K}_\infty  \right)}{\sigma\left(\int_{\infty}^{P'} \mathrm{d}\boldsymbol{u} + \int_{(e_2,0)}^{(Z,W)} \mathrm{d}\boldsymbol{u} - \boldsymbol{K}_\infty  \right)}.\end{split} \label{main1-2}
\end{align}
$P$ and $P'$ are points on the hyperelliptic curve, $Z$ is a pole, $W=w(Z)$ and $w^2=X(x)$. The zeros $e_i$ of $w^2(x)$ are the branch points of the curve $w^2$. $\mathrm{d}\boldsymbol{u}$ is the vector of the holomorphic differentials of the first kind $\mathrm{d}u_i=\frac{x^{i-1}}{w}\mathrm{d}x$ with $i=1,...,g$. $\zeta$ and $\sigma$ are Kleinian functions and $\boldsymbol{K}_\infty$ is the vector of Riemann constants. 

The vector $\boldsymbol{\mathfrak A}_i$ identified with each branch point $e_i$ is defined as \cite{Enolski:2011id}
\begin{equation}
\boldsymbol{\mathfrak{A}}_i=\int_{\infty}^{(e_i,0)} \mathrm{d}\boldsymbol{u}= 2\omega \boldsymbol{\varepsilon}_k+2\omega' \boldsymbol{\varepsilon}_i', \quad i=1,\ldots,6 \,,
\label{eqn:characteristics}
\end{equation}
with the vectors $\boldsymbol{\varepsilon}_i$ and $\boldsymbol{\varepsilon}_i'$ whose entries $\varepsilon_{i,j}$, $\varepsilon'_{i,j}$ are $\frac{1}{2}$ or $0$ for every $i=1,\ldots,6$, $j=1,2$. The matrix
\begin{equation}
 [\boldsymbol{u}_i] = \left[
 \begin{array}{c}
  \boldsymbol{\varepsilon}_i'\\
  \boldsymbol{\varepsilon}_i
 \end{array}
\right] 
\end{equation}
is called the characteristic of a branch point $e_i$.\\

The $g$th component (in this case genus $g=2$) of the vector $\boldsymbol{\mathfrak{Z}}(Z,W)$ is $\mathfrak{Z}_g(Z,W)=0$ and for $1\leq j<g$ we have
\begin{equation}
\mathfrak{Z}_j(Z,W)=\frac{W}{\prod_{k=2}^{g} (Z-e_{2k})}\sum_{k=0}^{g-j-1}(-1)^{g-k+j+1}Z^kS_{g-k-j-1}(\boldsymbol{e}) \, .
 \end{equation}
The $S_k(\boldsymbol{e})$ are elementary symmetric functions of order $k$ built on  $g-1$ branch points $e_4,\ldots, e_{2g}$: $S_0=1$, $S_1=e_4+\ldots+e_{2g}$, etc.\\

Then the solution of the $\phi$-equation reads
\begin{equation}
\begin{split}
 \phi &=\phi_{\rm in} + \Phi  \sum _{i=1}^2 K_i \left[ \frac{2}{W_i} \left(\int _{x_{\rm in}}^x d\boldsymbol{u}\right)^T \left( \boldsymbol{\zeta} \left( \int_{(e_2,0)}^{(p_i,W_i)} \mathrm{d} \boldsymbol{u} + \boldsymbol{K}_\infty  \right) - 2( \boldsymbol{\eta}^{\prime}\boldsymbol{\varepsilon}^\prime + \boldsymbol{\eta}\boldsymbol{\varepsilon} )  - \frac12 \boldsymbol{\mathfrak{Z}}(p_i,W_i)  \right) \right. \\ 
& \left. + \ln\frac{\sigma\left(  W^2(x)  \right)}{\sigma\left( W^1(x) \right)}
-  \ln \frac{\sigma\left(  W^2(x_{\rm in})  \right)}{\sigma\left( W^1(x_{\rm in}) \right)}  \right] 
\end{split}
\end{equation}
where $W_i=\sqrt{X(p_i)}$ and $W^{1,2}(x) = \int^{x}_{\infty}{d\boldsymbol{u}} \pm  \int_{(e_2,0)}^{(p_i,W_i)} \mathrm{d} \boldsymbol{u} - \boldsymbol{K}_\infty $.


\subsection{Solution of the $t$-equation}

With (\ref{eqn:sing-psi-x-gleichung}) equation (\ref{eqn:sing-psi-t-gleichung}) yields
\begin{equation}
\frac{\mathrm{d}t}{\mathrm{d}\gamma} =\frac{R^2EH^2(x)}{(x+1)(1-\lambda)^2}
\end{equation}
or
\begin{equation}
 t - t_{\rm in} = \frac{R^2E}{(1-\lambda )^2} \int _{x_{\rm in}}^x \! \frac{H^2(x')}{(x'+1)} \, \frac{\mathrm{d}x'}{\sqrt{X(x')}} \,.
\label{eqn:psi-tint}
\end{equation}
Next we apply a partial fractions decomposition upon (\ref{eqn:psi-tint}) where the constants $M_i$ arise.
\begin{equation}
 t - t_{\rm in} = \frac{R^2E}{(1-\lambda )^2} \int _{x_{\rm in}}^x \! \left( \frac{K_1}{x+1} + K_2 + K_3\cdot x \right) \, \frac{\mathrm{d}x'}{\sqrt{X(x')}}
\label{eqn:sing-t-pbz}
\end{equation}
First we will solve the holomorphic integrals $\int _{x_{\rm in}}^x\! \frac{x'\,^i}{\sqrt{X(x')}}\mathrm{d}x'$. 

We introduce a variable $v$ so that $v-v_{0} =  \int _{x_{\rm in}}^x\! \frac{\mathrm{d}x'}{\sqrt{X(x')}}$. The inversion of this integral yields $x(v)=-\frac{\sigma_{1}(\boldsymbol{u})}{\sigma_{2}(\boldsymbol{u})}$ (see section \ref{sec:sing-x-solution} or \cite{Enolski:2011id}), where
\begin{equation}
\boldsymbol{u}=\boldsymbol{\mathfrak A}_i+\left( 
\begin{array}{c} 
 v - v_0 \\ 
 f_1(v - v_0) 
 \end{array}   \right) ,\quad f_1(0)=0 \, .
\end{equation}
The function $f_1(v - v_0)$ can be found from the condition $\sigma(\boldsymbol{u})=0$.
Equation (\ref{eqn:sing-t-pbz}) now reads
\begin{equation}
 t - t_{\rm in} = \frac{R^2E}{(1-\lambda )^2} \left[ \int _{x_{\rm in}}^x \! \frac{K_1}{x+1} \, \frac{\mathrm{d}x'}{\sqrt{X(x')}} + K_2(v-v_0) + K_3  f_1(v-v_0) \right] \, .
\end{equation}
The remaining differential is of the third kind. Its solution is presented in (\ref{main1-2}). Then the solution of the $t$-equation (\ref{eqn:sing-psi-t-gleichung}) is
\begin{equation}
\begin{split}
 t &=t_{\rm in} + \frac{R^2E}{(1-\lambda )^2} \left\lbrace  K_1 \left[ \frac{2}{\sqrt{X(-1)}} \left(\int _{x_{\rm in}}^x d\boldsymbol{u}\right)^T \left( \boldsymbol{\zeta} \left( \int_{(e_2,0)}^{(-1,\sqrt{X(-1)})} \mathrm{d} \boldsymbol{u} + \boldsymbol{K}_\infty  \right) - 2( \boldsymbol{\eta}^{\prime}\boldsymbol{\varepsilon}^\prime + \boldsymbol{\eta}\boldsymbol{\varepsilon} )  \right.  \right. \right.\\ 
& \left. \left. \left. - \frac12 \boldsymbol{\mathfrak{Z}}(-1,\sqrt{X(-1)})  \right) + \ln\frac{\sigma\left(  W^2(x)  \right)}{\sigma\left( W^1(x) \right)}
-  \ln \frac{\sigma\left(  W^2(x_{\rm in})  \right)}{\sigma\left( W^1(x_{\rm in}) \right)}  \right] + K_2(v - v_0) + K_3f_1(v - v_0) \right\rbrace \\
\end{split}
\end{equation}
where $W^{1,2}(x) = \int^{x}_{\infty}{d\boldsymbol{u}} \pm  \int_{(e_2,0)}^{(-1,\sqrt{X(-1)})} \mathrm{d} \boldsymbol{u} - \boldsymbol{K}_\infty $.

\subsection{The orbits}
\label{sec:psi-axis-orbits}

\begin{figure}
 \centering
 \subfigure[$a$-$b$-plot ($x$-$y$-plane)\newline 
  The black dashed circles show the position of the horizon and the red dotted circles mark the ergosphere.]{
   \includegraphics[width=6cm]{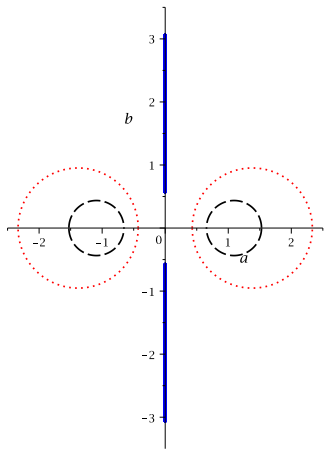}
 }
 \subfigure[Solution $x(\gamma)$\newline 
 The black horizontal lines are the position of the turning points.]{
   \includegraphics[width=6cm]{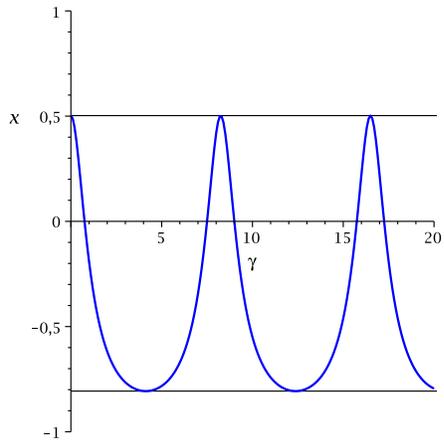}
 }
 \subfigure[$x_1$-$x_2$-plot ($x$-$\phi$-plane)]{
   \includegraphics[width=6cm]{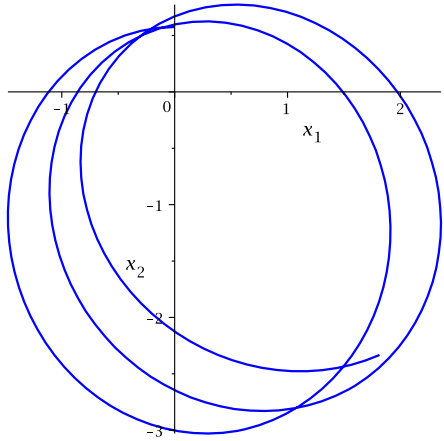}
 }
 \caption{$R=1$, $m=1$, $\lambda=0.4$, $\Phi=1$ and $E=0.9$: Bound orbit on the rotational axis.}
 \label{pic:sing-psi-bo}
\end{figure}

\begin{figure}
 \centering
 \subfigure[$a$-$b$-plot ($x$-$y$-plane)\newline 
  The black dashed circles show the position of the horizon and the red dotted circles mark the ergosphere.]{
   \includegraphics[width=6cm]{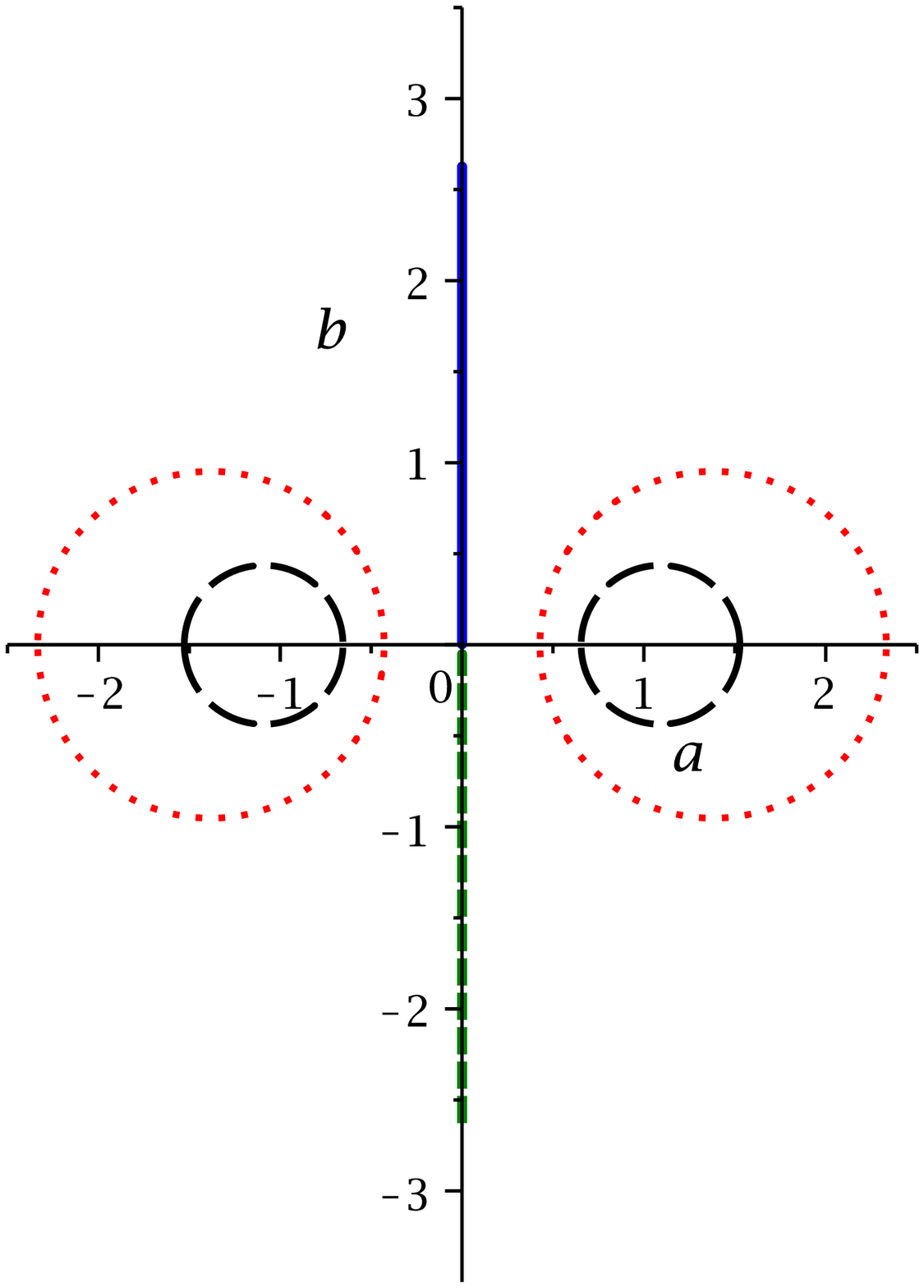}
 }
 \subfigure[Solution $x(\gamma)$\newline 
    The lower horizontal black line shows the turning point on each side of the ring. The upper horizontal black line at $x=1$ represents the equatorial plane, if a test particle reaches $x=1$ it continues its orbit on the other side of the ring.]{
   \includegraphics[width=6cm]{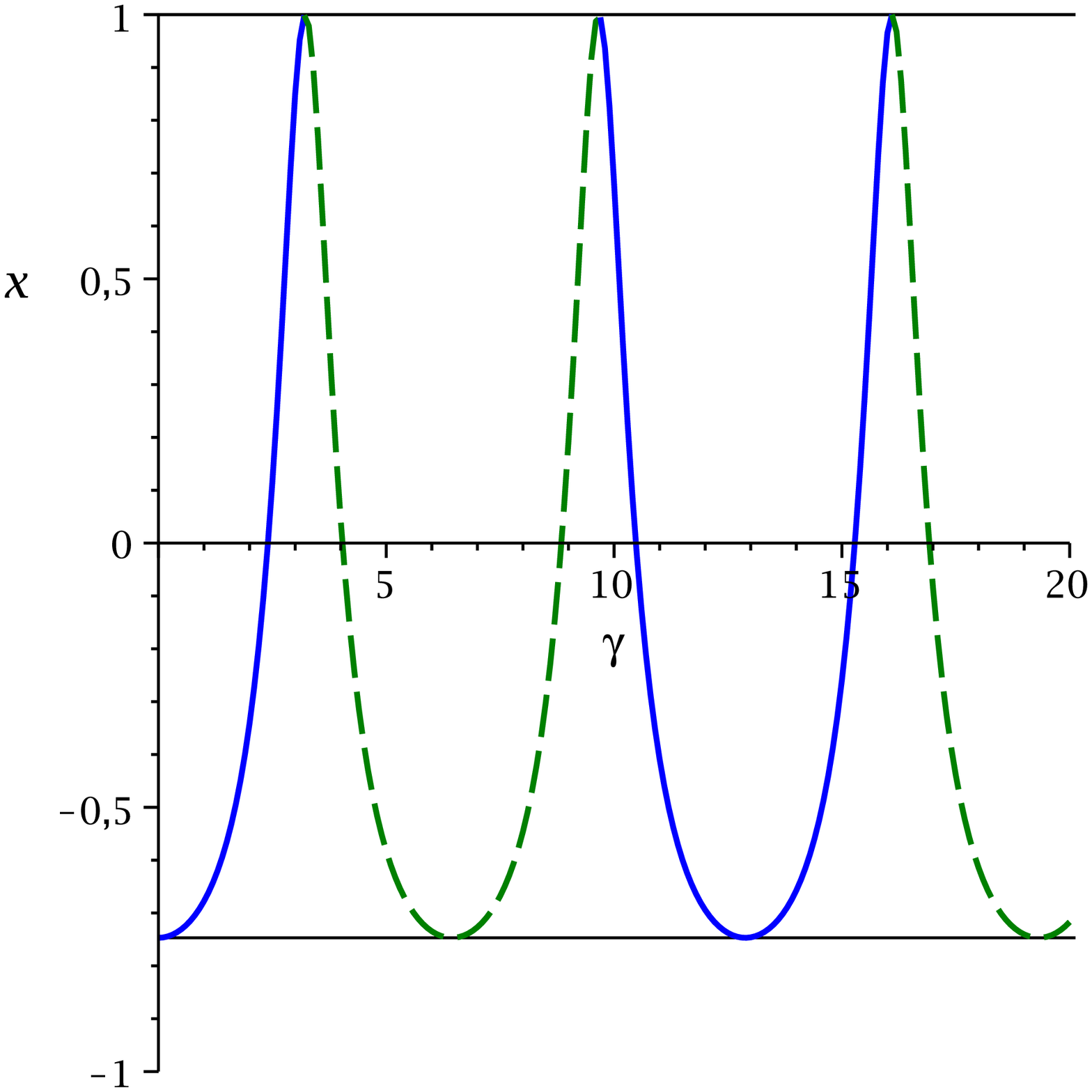}
 }
 \caption{$R=1$, $m=1$, $\lambda=0.4$, $\Phi=0$ and $E=0.8$: Bound orbit passing through the black ring on the rotational axis. The motion above the equatorial plane is shown in blue (solid) and the motion below the equatorial plane is shown in green (dashed).}
 \label{pic:sing-psi-bo2}
\end{figure}

\begin{figure}
 \centering
 \subfigure[$a$-$b$-plot ($x$-$y$-plane)\newline 
  The black dashed circles show the position of the horizon and the red dotted circles mark the ergosphere.]{
   \includegraphics[width=6cm]{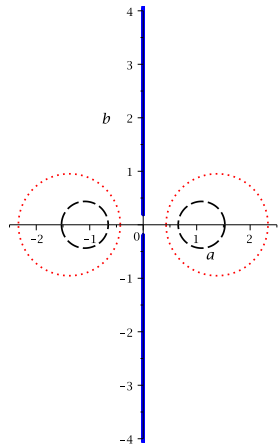}
 }
 \subfigure[Solution $x(\gamma)$\newline 
 Solution $x(\gamma)$\newline 
    The upper horizontal black line is the position of the turning point. The lower horizontal black line represents infinity ($y=x=-1$ in ring coordinates).]{
   \includegraphics[width=6cm]{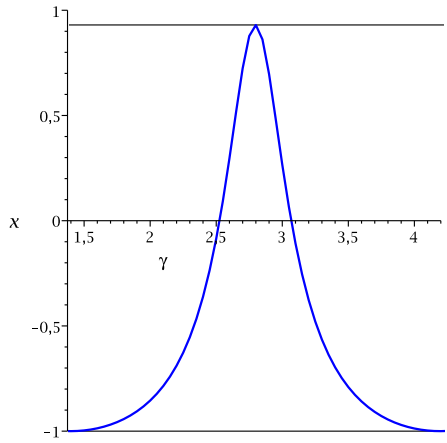}
 }
 \subfigure[$x_1$-$x_2$-plot ($x$-$\phi$-plane)]{
  \includegraphics[width=6cm]{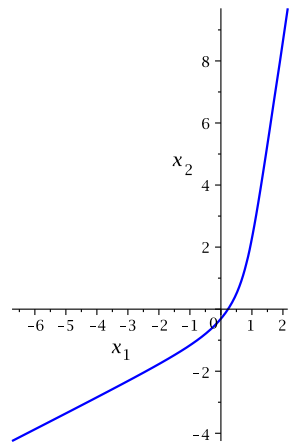}
 }
 \caption{$R=1$, $m=1$, $\lambda=0.4$, $\Phi=1$ and $E=2$: Escape orbit on the rotational axis.}
 \label{pic:sing-psi-eo}
\end{figure}

\begin{figure}
 \centering
 \subfigure[$a$-$b$-plot ($x$-$y$-plane)\newline 
  The black dashed circles show the position of the horizon and the red dotted circles mark the ergosphere.]{
   \includegraphics[width=6cm]{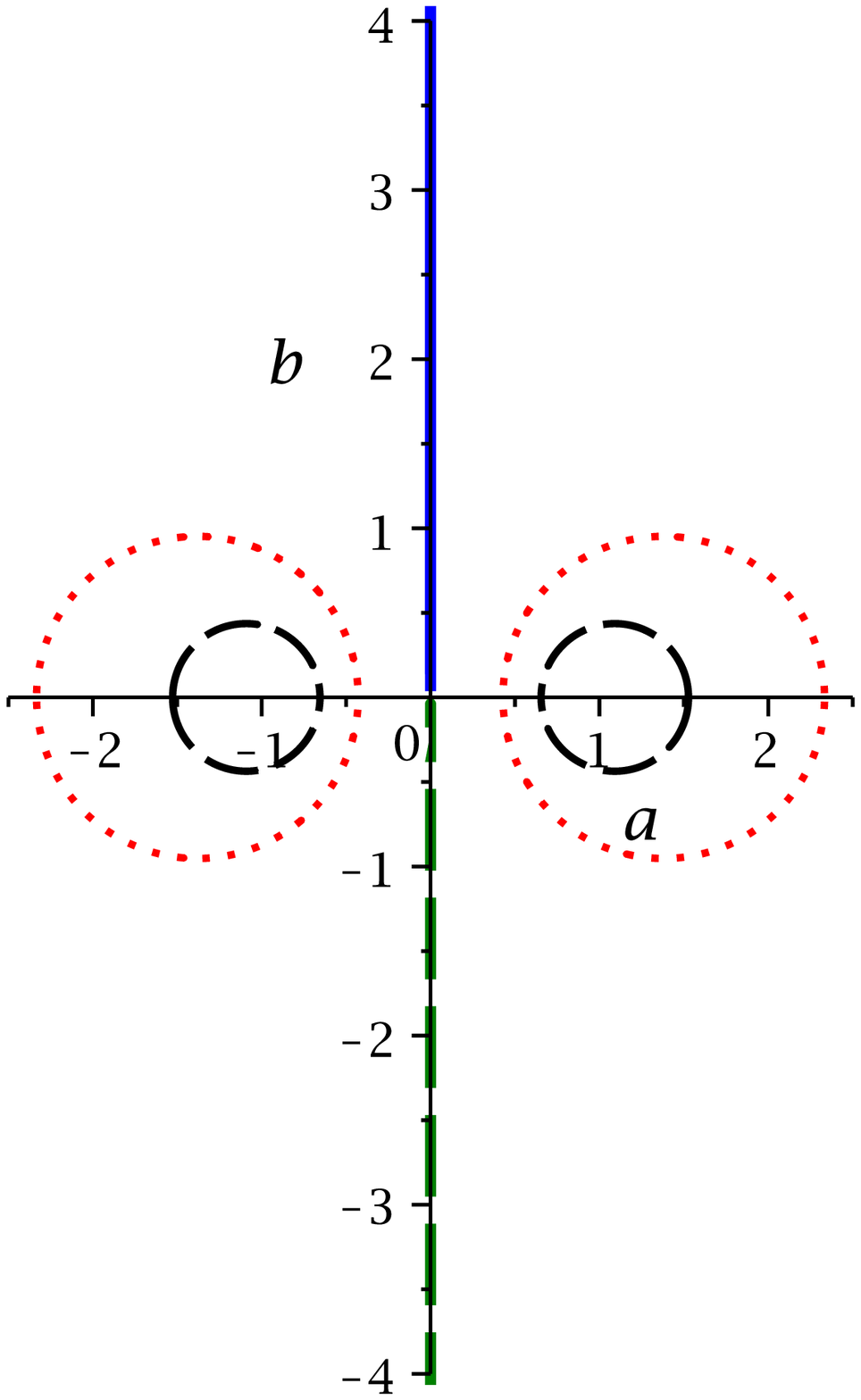}
 }
 \subfigure[Solution $x(\gamma)$\newline 
    The upper horizontal black line at $x=1$ represents the equatorial plane, if a test particle reaches $x=1$ it continues its orbit on the other side of the ring. The lower horizontal black line represents infinity ($y=x=-1$ in ring coordinates).]{
   \includegraphics[width=6cm]{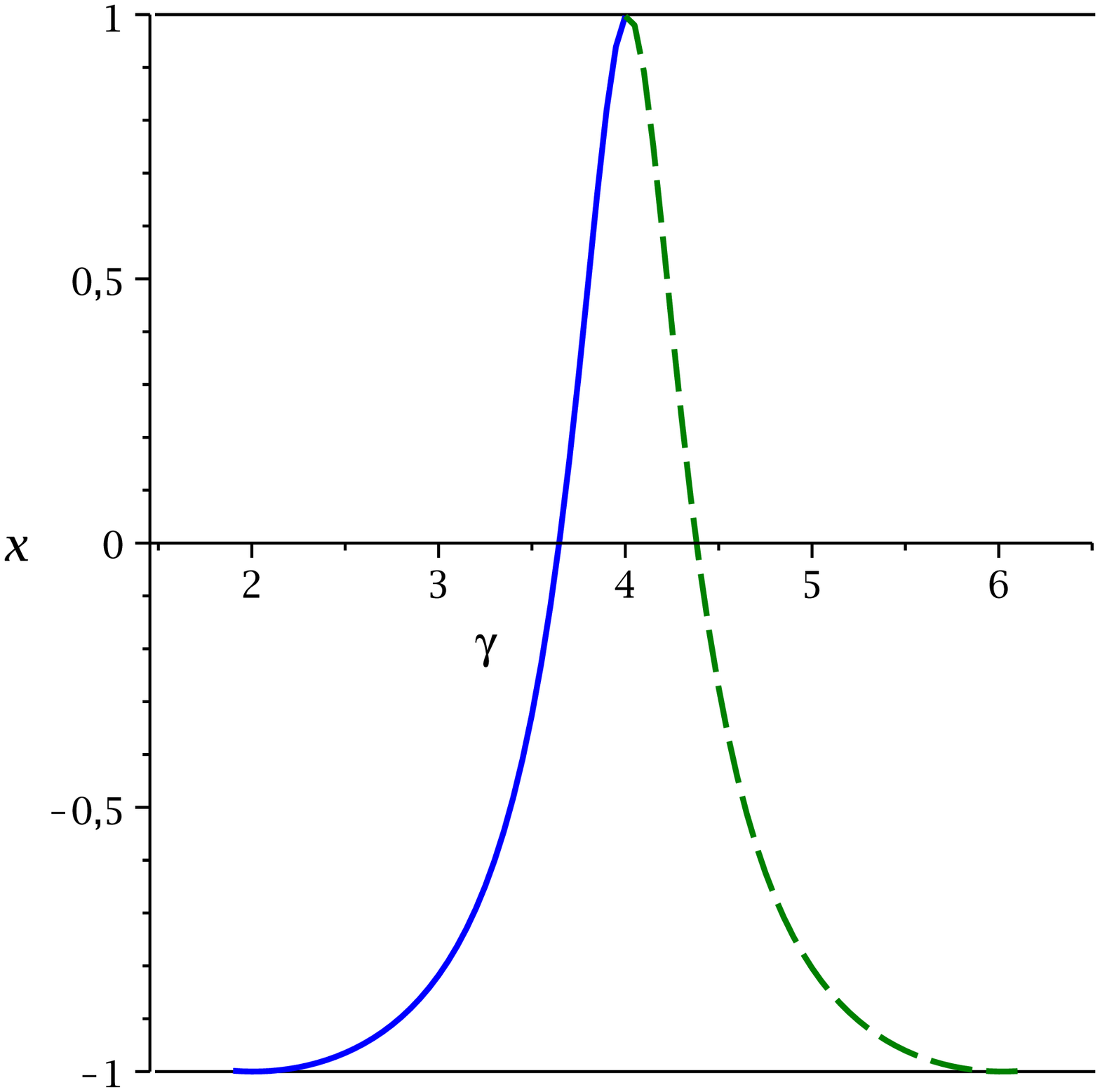}
 }
 \caption{$R=1$, $m=1$, $\lambda=0.4$, $\Phi=0$ and $E=1.5$: Escape orbit passing through the black ring on the rotational axis. The motion above the equatorial plane is shown in blue (solid) and the motion below the equatorial plane is shown in green (dashed).}
 \label{pic:sing-psi-eo2}
\end{figure}

On the rotational axis of the singly spinning black ring bound orbits and escape orbits are possible. The orbits either move around the ring or move directly through the center of the ring. Figure \ref{pic:sing-psi-bo} and \ref{pic:sing-psi-bo2} show bound orbits and figure \ref{pic:sing-psi-eo} and  \ref{pic:sing-psi-eo2} show escape orbits in $a$-$b$-coordinates (see section \ref{sec:em0-orbit}) and the corresponding solution $x(\gamma)$. Since the orbits on the rotational axis are presented by lines in the $x$-$y$-plane, we will also show them in the $x$-$\phi$-plane where we will use the coordinates $r_1$ and $\phi$.

One can think of ring coordinates as two pairs of polar coordinates
\begin{equation}
\begin{array}{l} 
x_1=r_1 \sin(\phi)\\
x_2=r_1 \cos(\phi)
\end{array}
\quad\text{and}\quad
\begin{array}{l} 
x_3=r_2 \sin(\psi)\\
x_4=r_2 \cos(\psi)
\end{array}
\end{equation}
where
\begin{equation}
 r_1=R\frac{\sqrt{1-x^2}}{x-y} \quad\text{and}\quad r_2=R\frac{\sqrt{y^2-1}}{x-y}
\end{equation}
(see \cite{Hoskisson:2007zk,Emparan:2006mm}).

If $\psi$ is constant, the horizon of the black ring consists of two $S^2$ spheres. If we look at the rotational axis where $y=-1$, the coordinates $x_1$ and $x_2$ describe the plane between these two spheres, so the horizon cannot be seen in this plane.
If $\phi$ is constant, the horizon has $S^1\times S^1$ topology. So if $x=\pm1$ the coordinates $x_3$ and $x_4$ describe the equatorial plane ``as seen from above''.\\

The bound orbits of type B and the escape orbits of type A, which move through the center of the black ring, are lines in every plane since here both angles $\phi$ and $\psi$ are constant. So in that case we only show the $a$-$b$-plot.

\section{Geodesics on the equatorial plane}

The surface $x=\pm 1$ is the equatorial plane of the black ring, which is divided into two parts. The first part $x=+1$ is the plane enclosed by the ring (or more precisely: enclosed by the singularity), which we will refer to as ``inside'' the ring. The second part $x=-1$ describes the equatorial plane around the black ring (or more precisely: around the singularity), which we will refer to as ``outside'' the ring.

If we set $x=\pm 1$, $\Phi =0$ and $p_x=\frac{\partial S}{\partial x}=0$ in the Hamilton-Jacobi equation (\ref{eqn:hjd-sring}), it depends on the coordinate $y$ only:
\begin{equation}
 0 = m^2 - \frac{(1\pm\lambda)^2}{H(y)}E^2 + \frac{(\pm 1-y)^2}{R^2(1\pm\lambda)^2}\left[-G(y)\left( \frac{\partial S}{\partial y}\right) ^2 - \frac{H(y)}{G(y)}(\Psi + \Omega_\psi E)^2 \right] \, .
\end{equation}
This can be rearranged to
\begin{equation}
 \left( \frac{\partial S}{\partial y}\right) ^2 = \frac{R^2(1\pm\lambda)^2}{(\pm 1-y)^2G(y)}\left[ m^2 - \frac{H(y)}{(1\pm\lambda)^2}E^2\right] - \frac{H(y)}{G^2(y)}(\Psi + \Omega_\psi E)^2 := Y_S \, .
\end{equation}
Then we have
\begin{equation}
 S=\frac{1}{2}m^2\tau -Et+\Psi\psi + \int\! \sqrt{Y_S} \, \mathrm{d}y \, .
\end{equation}
Now we set the derivatives of $S$ with respect to the constants $m^2$, $E$ and $\Psi$ to zero in order to obtain the equations of motion.
With the Mino-time \cite{Mino:2003yg} $\mathrm{d}\gamma=\frac{\pm 1-y}{R^2}\mathrm{d}\tau$ and the relation  $\Omega_\psi = -CR\frac{1+y}{H(y)}$ the equations of motion take the form
\begin{eqnarray}
 \frac{\mathrm{d}y}{\mathrm{d}\gamma} &=& \left\lbrace R^2\frac{G(y)}{H(y)}\left[ \frac{H(y)}{(1\pm\lambda)^2}m^2-E^2\right] - \frac{(\pm 1-y)^2H(y)}{(1\pm\lambda)^4} [H(y)\Psi +\Omega_\psi]^2 \right\rbrace ^{1/2}  \nonumber\\
      &:=& \sqrt{Y(y)} \, , \label{eqn:sing-phi-y-gleichung}\\
 \frac{\mathrm{d}\psi}{\mathrm{d}\gamma} &=& -\frac{(\Psi + \Omega_\psi E)(\pm 1-y)H(y)}{(1\pm\lambda)^2G(y)} \, ,\label{eqn:sing-phi-psi-gleichung}\\
 \frac{\mathrm{d}t}{\mathrm{d}\gamma} &=& \frac{R^2E}{(\pm 1-y)H(y)} + \frac{(\Omega_\psi\Psi+\Omega_\psi^2E)(\pm 1-y)H(y)}{(1\pm\lambda)^2G(y)} \label{eqn:sing-phi-t-gleichung} \, .
\end{eqnarray}
It might not be obvious at first glance, but $Y(y)$ is a polynomial of third order in $y$ and therefore the equations of motion are of elliptic type.

\subsection{Classification of geodesics}

From (\ref{eqn:sing-phi-y-gleichung}) we can read off an effective potential consisting of the two parts $V_+(y)$ and $V_-(y)$:
\begin{equation}
 Y=a(y)(E-V_+)(E-V_-) \, .
\end{equation}
Since $Y(y)$ can be written as $Y(y)=a(y)E^2+b(y)E+c(y)$ the effective potential takes the form
\begin{equation}
 V_\pm (y) = \frac{-b(y)\pm\sqrt{b(y)^2-4a(y)c(y)}}{2a(y)}, \qquad \mathrm{where}
\end{equation}
\begin{eqnarray}
 a(y) &=& -R^2\frac{G(y)}{H(y)}-\frac{(\pm1-y)^2C^2R^2(1+y)^2}{(1\pm\lambda)^4H(y)}\, , \nonumber\\
 b(y) &=& \frac{2(\pm1-y)^2\Psi CR(1+y)}{(1\pm\lambda)^4}\, ,\nonumber \\
 c(y) &=& \frac{R^2G(y)m^2}{(1\pm\lambda)^2}-\frac{(\pm 1-y)^2H(y)\Psi^2}{(1\pm\lambda)^4} \, .
\end{eqnarray}
The two cases $x=+1$ (geodesics inside the ring) and $x=-1$ (geodesics outside the ring) have to be discussed separately.

\subsubsection{Geodesics outside the ring}

\begin{figure}
 \centering
 \subfigure[$\Psi=0$ \newline Examples of orbits of type A, B and B$_0$. The potential is symmetric and $Y(y)$ has none or one zeros.]{
   \includegraphics[width=7cm]{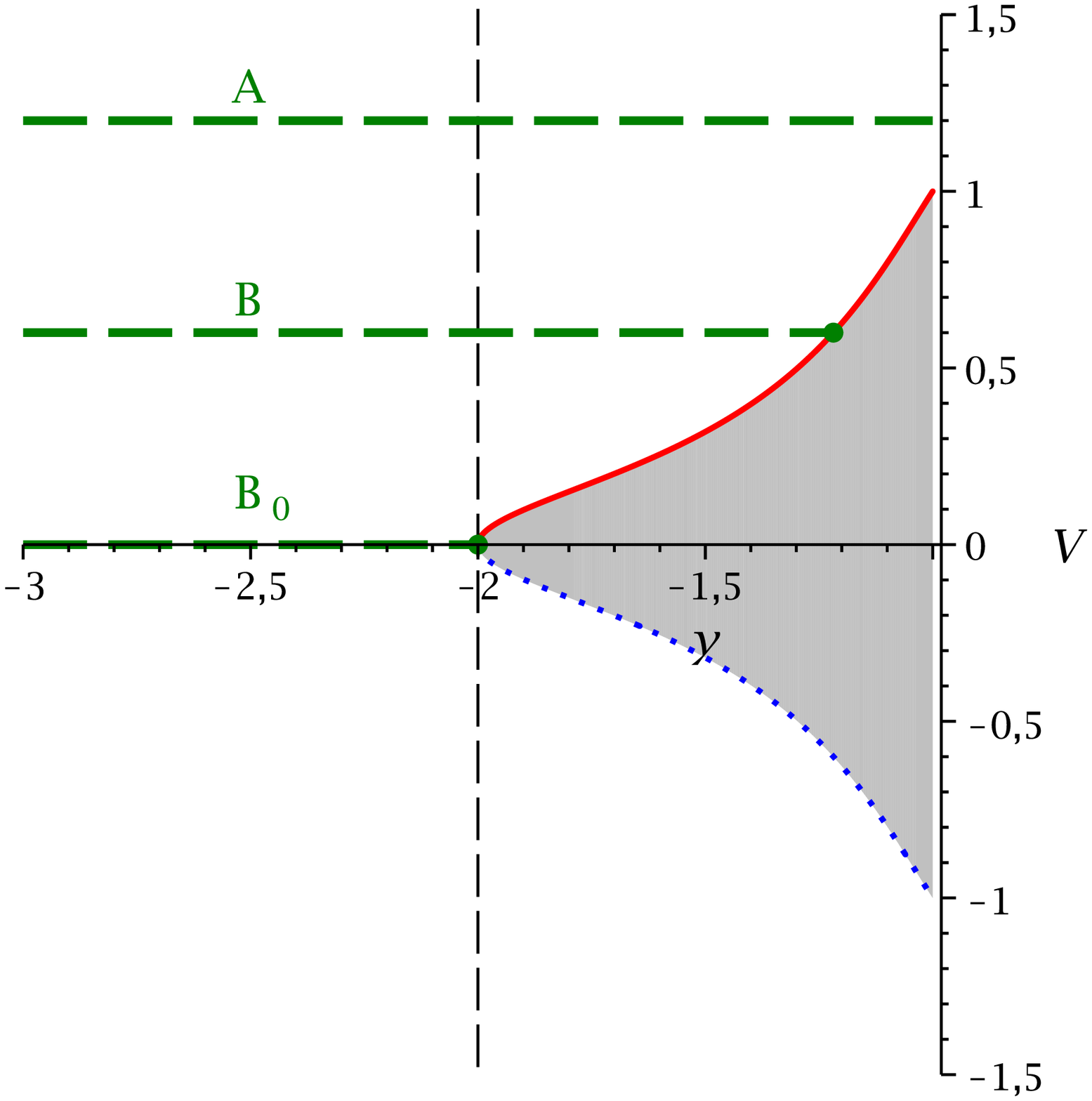}
 }
 \subfigure[$\Psi=5$ \newline Examples of orbits of type of type C and C$_0$. If $|\Psi|>0$ the potential is no longer symmetric and up to two zeros of $Y(y)$ are possible.]{
   \includegraphics[width=7cm]{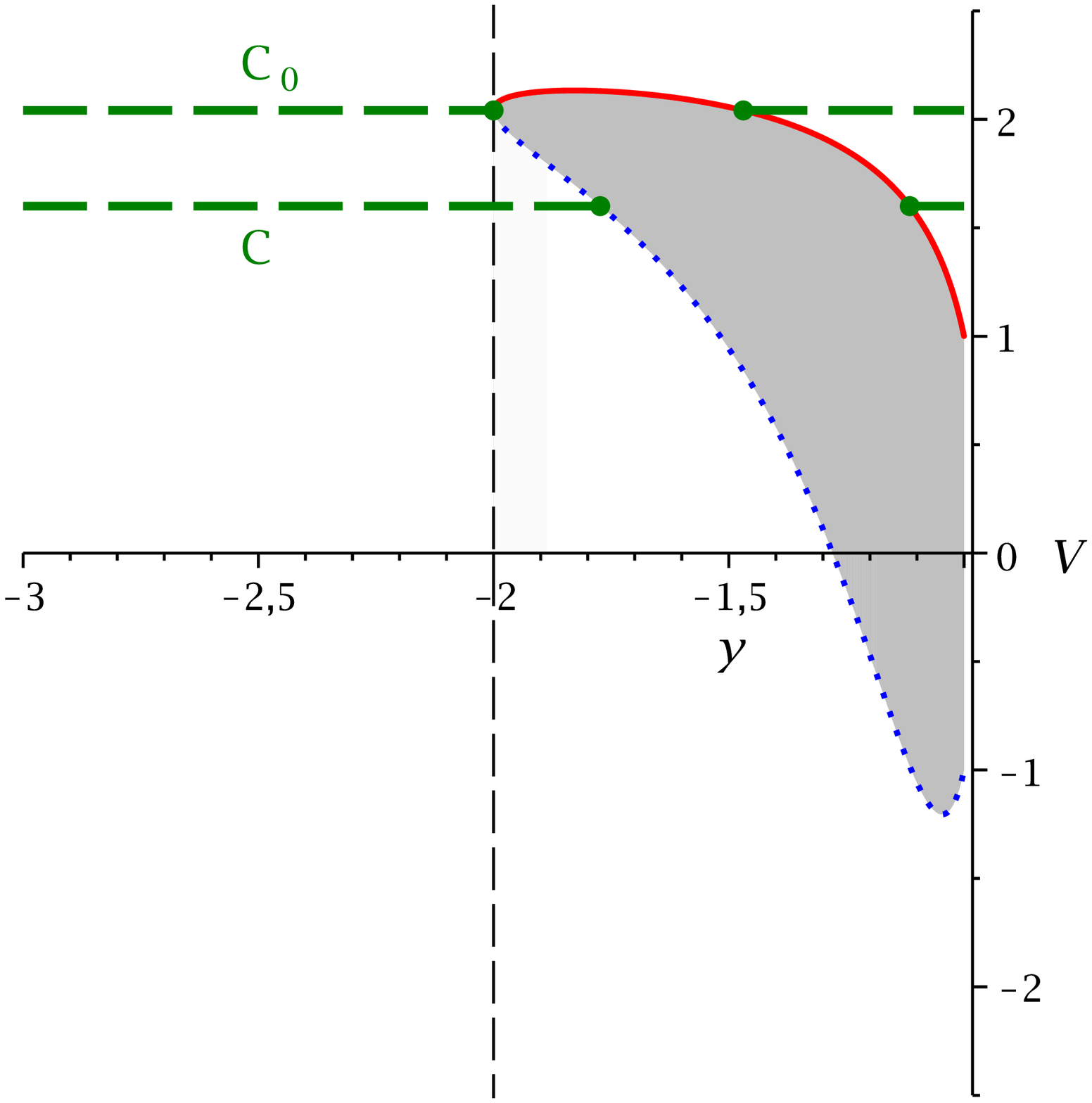}
 }
 \caption{$R=1$, $m=1$ and $\lambda=0.5$ \newline
    Effective potentials $V_+(y)$ (red, solid) and $V_-(y)$ (blue, dotted) on the $\phi$ axis outside the ring. The grey area is a forbidden zone, where no motion is possible. The horizon is marked by a vertical dashed line. Green dashed lines represent energys and green points mark the turning points.}
 \label{pic:sing-phi-orbits1}
\end{figure}

Let us first take a look at the motion on the surface outside the black ring. Here we have $x=-1$. Figure \ref{pic:sing-phi-orbits1} shows the effective potential $V(y)$ for different values of the parameters. $V_+$ and $V_-$ meet at the horizon. Mainly the angular momentum $\Psi$ defines the shape of the effective potential. For $\Psi =0$ the potential is symmetric and $Y(y)$ has none or one zero. If $|\Psi|>0$ the potential is no longer symmetric and if $|\Psi|$ is large enough up to two zeros of $Y(y)$ are possible.

Possible orbits are Terminating Orbits (TO) with or without a turning point, where light or test particles cross the horizon and fall into the singularity, and Escape Orbits (EO), where light or test particles aproach the black ring, turn around at a certain point and escape the gravitational field. The zero of $Y(y)$ of a TO can lie directly on the event horizon.

There are three different types of orbits (see table \ref{tab:sing-phi-typen-orbits1}).

\begin{itemize}
 \item Type A:\\
  $Y(y)$ has no zeros and only TOs exist.
 \item Type B:\\
  $Y(y)$ has one zero and only TOs exist. In a special case the zero of $Y(y)$ lies on the horizon.
 \item Type C:\\
  $Y(y)$ has two zeros. TOs and EOs exist. In a special case the zero of $Y(y)$ lies on the horizon.
\end{itemize}

\begin{table}[ht]
\begin{center}
\begin{tabular}{|lcll|}\hline
type &  zeros  & range of $y$ & orbit \\
\hline\hline
A & 0 & 
\begin{pspicture}(-2.5,-0.2)(3,0.2)
\psline[linewidth=0.5pt]{|->}(-2.5,0)(3,0)
\psline[linewidth=0.5pt,doubleline=true](0.5,-0.2)(0.5,0.2)
\psline[linewidth=1.2pt]{-}(-2.5,0)(3,0)
\end{pspicture}
  & TO
\\  \hline
B  & 1 &
\begin{pspicture}(-2.5,-0.2)(3,0.2)
\psline[linewidth=0.5pt]{|->}(-2.5,0)(3,0)
\psline[linewidth=0.5pt,doubleline=true](0.5,-0.2)(0.5,0.2)
\psline[linewidth=1.2pt]{-*}(-2.5,0)(1,0)
\end{pspicture}
& TO 
\\ \hline
B$_0$ & 1 & 
\begin{pspicture}(-2.5,-0.2)(3,0.2)
\psline[linewidth=0.5pt]{|->}(-2.5,0)(3,0)
\psline[linewidth=0.5pt,doubleline=true](0.5,-0.2)(0.5,0.2)
\psline[linewidth=1.2pt]{-*}(-2.5,0)(0.5,0)
\end{pspicture}
  & TO 
\\ \hline
C & 2 & 
\begin{pspicture}(-2.5,-0.2)(3,0.2)
\psline[linewidth=0.5pt]{|->}(-2.5,0)(3,0)
\psline[linewidth=0.5pt,doubleline=true](0.5,-0.2)(0.5,0.2)
\psline[linewidth=1.2pt]{-*}(-2.5,0)(1,0)
\psline[linewidth=1.2pt]{*-}(2.0,0)(3,0)
\end{pspicture}
  & TO, EO
\\ \hline
C$_0$ & 2 & 
\begin{pspicture}(-2.5,-0.2)(3,0.2)
\psline[linewidth=0.5pt]{|->}(-2.5,0)(3,0)
\psline[linewidth=0.5pt,doubleline=true](0.5,-0.2)(0.5,0.2)
\psline[linewidth=1.2pt]{-*}(-2.5,0)(0.5,0)
\psline[linewidth=1.2pt]{*-}(2.0,0)(3,0)
\end{pspicture}
& TO, EO 
\\ \hline\hline
\end{tabular}
\caption{Types of orbits of light and particles in the singly spinning black ring spacetime for $x=-1$, $\Phi =0$. The thick lines represent the range of the orbits. The turning points are shown by thick dots. The horizon is indicated by a vertical double line. In a special case the zero of $Y(y)$ lies on the horizon.}
\label{tab:sing-phi-typen-orbits1}
\end{center}
\end{table}

\subsubsection{Geodesics inside the ring}

\begin{figure}[ht]
 \centering
 \includegraphics[width=6cm]{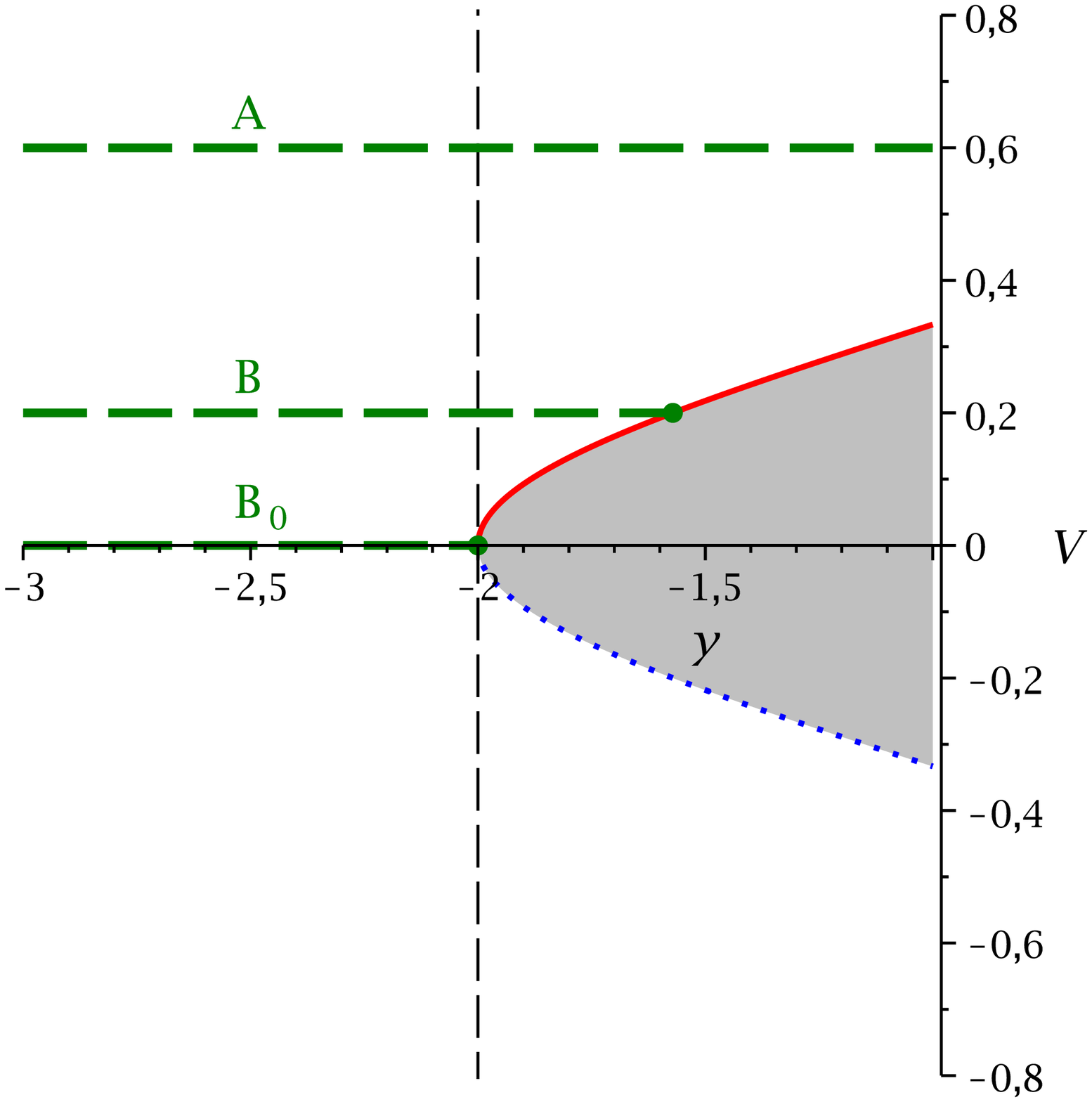}
 \caption{$R=1$, $m=1$, $\lambda=0.5$ und $\Phi=0$ \newline
    Effective potentials $V_+(y)$ (red, solid) and $V_-(y)$ (blue, dotted) on the equatorial plane inside the ring. The grey area is a forbidden zone, where no motion is possible. The horizon is marked by a vertical dashed line. Green dashed lines represent energies and green points mark the turning points. Possible orbits are of type A, B and B$_0$.}
 \label{pic:sing-phi-orbits2}
\end{figure}

The effective potential for geodesics on the surface enclosed by the black ring ($x=+1$) is shown in figure \ref{pic:sing-phi-orbits2}. Again, if we have $\Psi =0$ the potential is symmetric and $Y(y)$ has none or one zeros. In the case $|\Psi|>0$ a potential barrier appears which prevents test particles and light from reaching $y=-1$. Then $Y(y)$ has always a single zero in the allowed range of $y$. The higher $|\Psi|$ the higher the energy where $V_+$ and $V_-$ meet. Note that $x=+1$, $y=-1$ is the location of the center of the black ring.

Possible orbits are Terminating Orbits where $Y(y)$ has one zero (type B) or none zero (type A). For the type of orbits see previous section. In a special case the zero of $Y(y)$ lies on the horizon. (type B$_0$).

\subsection{Solution of the $y$-equation}
\label{sec:sing-phi-y-solution}

Equation (\ref{eqn:sing-phi-y-gleichung}) can be solved analogously to (\ref{eqn:sing-x-gleichung}). (\ref{eqn:sing-phi-y-gleichung}) can be written as
\begin{equation}
 \left( \frac{\mathrm{d}y}{\mathrm{d}\gamma}\right)^2 = Y(y) = b_3 y^3 +b_2y^2+b_1y+b_0 \, ,
\end{equation}
where
\begin{equation}
 \begin{split}
  b_3&=\frac{-C^2R^2E^2}{2\lambda(\lambda\pm1)^4} + \frac{CRE\Psi-\lambda\Psi^2}{(\lambda\pm1)^4} - \frac{R^2m^2\lambda}{(\lambda\pm1)^2}\\
  b_2&=\frac{C^2R^2E^2}{2\lambda^2(\lambda\pm1)^4} + \frac{4(1\pm1)\lambda(1+\lambda+2\lambda^2)R^2E^2}{(\lambda-1)(\lambda\pm1)^4} + \frac{(-\lambda^2\pm4\lambda-1)\Psi^2}{(\lambda\pm1)^4}\mp\frac{2(2\mp1)CRE\Psi}{(\lambda\pm1)^4}-\frac{R^2m^2}{(\lambda\pm1)^2}\\
  b_1&=\frac{12\lambda^2R^2E^2}{(\lambda-1)(\lambda\pm1)^4}-\frac{\lambda(\lambda+1)R^2E^2}{(\lambda-1)(\lambda\pm1)^2}\mp\frac{2(2\mp1)CRE\Psi}{(\lambda\pm1)^4} \pm \frac{2(\lambda^2\mp\lambda+1)\Psi^2}{(\lambda\pm1)^4} + \frac{R^2m^2\lambda}{(\lambda\pm1)^2}\\
  b_0&= \frac{-C^2R^2E^2}{2\lambda^2(\lambda\pm1)^4}  + \frac{2CRE\Psi-(\lambda^2+1)\Psi^2}{(\lambda\pm1)^4} + \frac{4(1\pm1)\lambda R^2E^2}{(\lambda\pm1)^4} + \frac{R^2m^2}{(\lambda\pm1)^2} \, .
 \end{split}
\end{equation}
The solution is (see section \ref{sec:ergo-xsol})
\begin{equation}
y (\gamma)=\frac{1}{b_{3}}\left[ 4\wp (\gamma - \gamma '_{\rm in},g_{2},g_{3}) -\frac{b_{2}}{3} \right] ,
\end{equation}
where  $\gamma'_{\rm in} = \gamma _{\rm in} + \int _{v_{\rm in}}^\infty \! \frac{\mathrm{d}v'}{\sqrt{4v'^3 - g_{2} v' -g_{3}}} $ and $v_{\rm in}=\frac{1}{4} \left( b_{3}y_{\rm in}+\frac{b_{2}}{3}\right) $. The coefficients $g_2$ and $g_3$ of the polynomial in the Weierstra{\ss} form are
\begin{equation}
g_{2}=\frac{b_{2}^2}{12} - \frac{b_{1} b_{3}}{4} \qquad \mathrm{and} \qquad
g_{3}=\frac{b_{1} b_{2} b_{3}}{48}-\frac{b_{0} b_{3}^2}{16}-\frac{b_{2}^3}{216} \ .
\end{equation}


\subsection{Solution of the $\psi$-equation}
\label{sec:sing-phi-psi-solution}

With (\ref{eqn:sing-phi-y-gleichung}) equation (\ref{eqn:sing-phi-psi-gleichung}) yields
\begin{equation}
     \mathrm{d}\psi = -\frac{(\Psi+\Omega_\psi E)(\pm1-y)H(y)}{(1\pm\lambda)^2G(y)} \frac{\mathrm{d}y}{\sqrt{Y(y)}}
\end{equation}
or
\begin{equation}
 \psi - \psi_{\rm in} = \int _{y_{\rm in}}^y \! -\frac{(\Psi+\Omega_\psi E)(\pm1-y')H(y')}{(1\pm\lambda)^2G(y')} \, \frac{\mathrm{d}y'}{\sqrt{Y(y')}} \, .
\end{equation}
This can be rewritten as
\begin{equation}
 \psi - \psi_{\rm in} = \frac{\pm1}{(1\pm\lambda)^2}\int _{y_{\rm in}}^y \! \frac{CRE(1+y')-H(y')\Psi}{(1\pm y)(1+\lambda y')} \, \frac{\mathrm{d}y'}{\sqrt{Y(y')}} \, .
\end{equation}
This equation can be solved analogously to the $\phi$- and $\psi$-equation for nullgeodesics in the ergosphere (see section \ref{sec:ergo-phisol} and \ref{sec:ergo-psisol}). With $v=v(\gamma)=\gamma-\gamma'_{\rm in}$, $v_{\rm in}=v(\gamma_{\rm in})$ and $p_j=\wp(v_j)$ the solution is
\begin{equation}
\begin{split}
 \psi (\gamma) &= \sum^2_{j=1} \frac{K_j}{\wp^\prime_y(v_{j})}\Biggl( 2\zeta_y(v_{j})(v-v_{\rm in}) + \log\frac{\sigma_y(v-v_{j})}{\sigma_y(v_{\rm in}-v_{j})} - \log\frac{\sigma_y(v+v_{j})}{\sigma_y(v_{\rm in}+v_{j})} \Biggr) \\
& + \psi _{\rm in} \, .
\end{split}
\end{equation}
$K_j$ are constants which arise from the partial fractions decomposition and depend on the parameters of the metric and the test particle.

\subsection{Solution of the $t$-equation}

With (\ref{eqn:sing-phi-y-gleichung}) equation (\ref{eqn:sing-phi-t-gleichung}) yields
\begin{equation}
     \mathrm{d}t = \left( \frac{R^2E}{H(y)(\pm1-y)} + \frac{(\pm1-y)CR[\Psi H(y)+ CRE(1+y)]}{(1\pm\lambda)^2(1-y)(1+\lambda y)H(y)} \right) \frac{\mathrm{d}y}{\sqrt{Y(y)}}
\end{equation}
or
\begin{equation}
     t - t_{\rm in} = \int _{y_{\rm in}}^y \! \left( \frac{R^2E}{H(y')(\pm1-y')} + \frac{(\pm1-y')CR[\Psi H(y')+ CRE(1+y')]}{(1\pm\lambda)^2(1-y')(1+\lambda y')H(y')} \right)\, \frac{\mathrm{d}y'}{\sqrt{Y(y')}} \, .
     \label{eqn:phi-tint}
\end{equation}
This equation can be solved analogously to the $\phi$- and $\psi$-equation for nullgeodesics in the ergosphere (see section \ref{sec:ergo-phisol} and \ref{sec:ergo-psisol}). With $v=v(\gamma)=\gamma-\gamma'_{\rm in}$, $v_{\rm in}=v(\gamma_{\rm in})$ and $q_j=\wp(v_j)$ the solution is
\begin{equation}
     \begin{split}
     t (\gamma) &= \sum^4_{j=1} \frac{M_j}{\wp^\prime_y(v_{j})}\Biggl( 2\zeta_y(v_{j})(v-v_{\rm in}) + \log\frac{\sigma_y(v-v_{j})}{\sigma_y(v_{\rm in}-v_{j})} - \log\frac{\sigma_y(v+v_{j})}{\sigma_y(v_{\rm in}+v_{j})} \Biggr)\\
     & + M_0(v-v_{\rm in}) + t _{\rm in} \, .
     \end{split}
\end{equation}
$M_j$ are constants which arise from the partial fractions decomposition and depend on the parameters of the metric and the test particle.

\subsection{The orbits}

\begin{figure}[h]
 \centering
 \subfigure[$a$-$b$-plot\newline 
  The black dashed circles show the position of the horizon and the red dotted circles mark the ergosphere.]{
   \includegraphics[width=9cm]{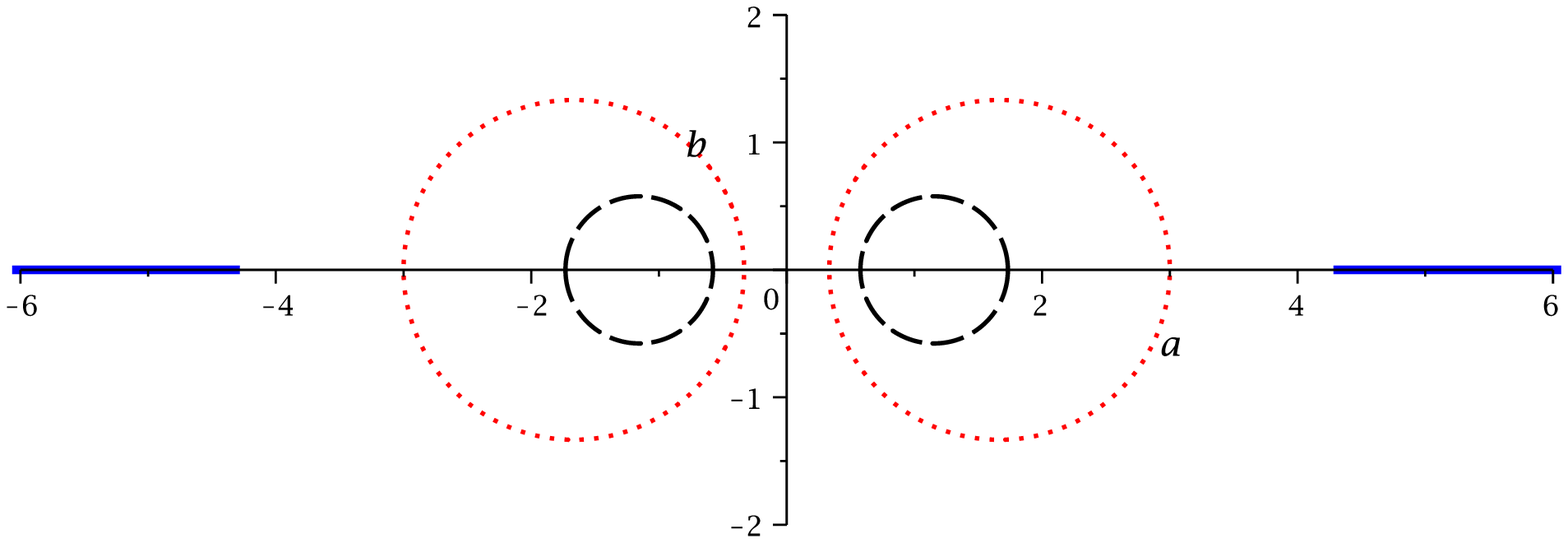}
 }
 \subfigure[Solution $y(\gamma)$\newline 
    The lower horizontal black line marks the position of the turning point and the upper horizontal black line shows where infinity is reached ($x=y=-1$ in ring coordinates)]{
   \includegraphics[width=6cm]{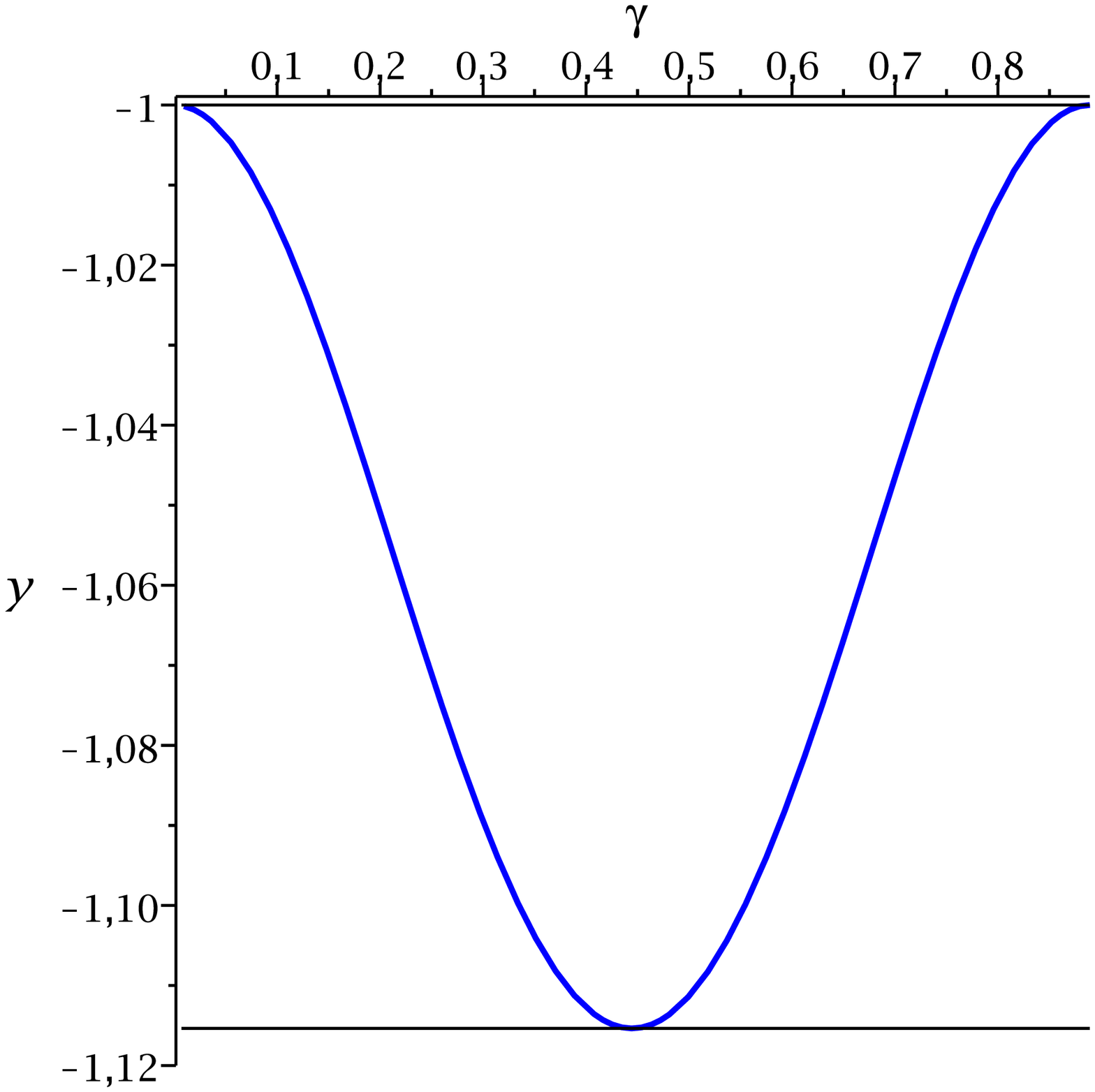}
 }
 \subfigure[$x_3$-$x_4$-plot ($y$-$\psi$-plane)\newline
 In this plane we are looking at the black ring from above. The black dashed circles show the position of the horizon and the red dotted circles mark the ergosphere.]{
  \includegraphics[width=6cm]{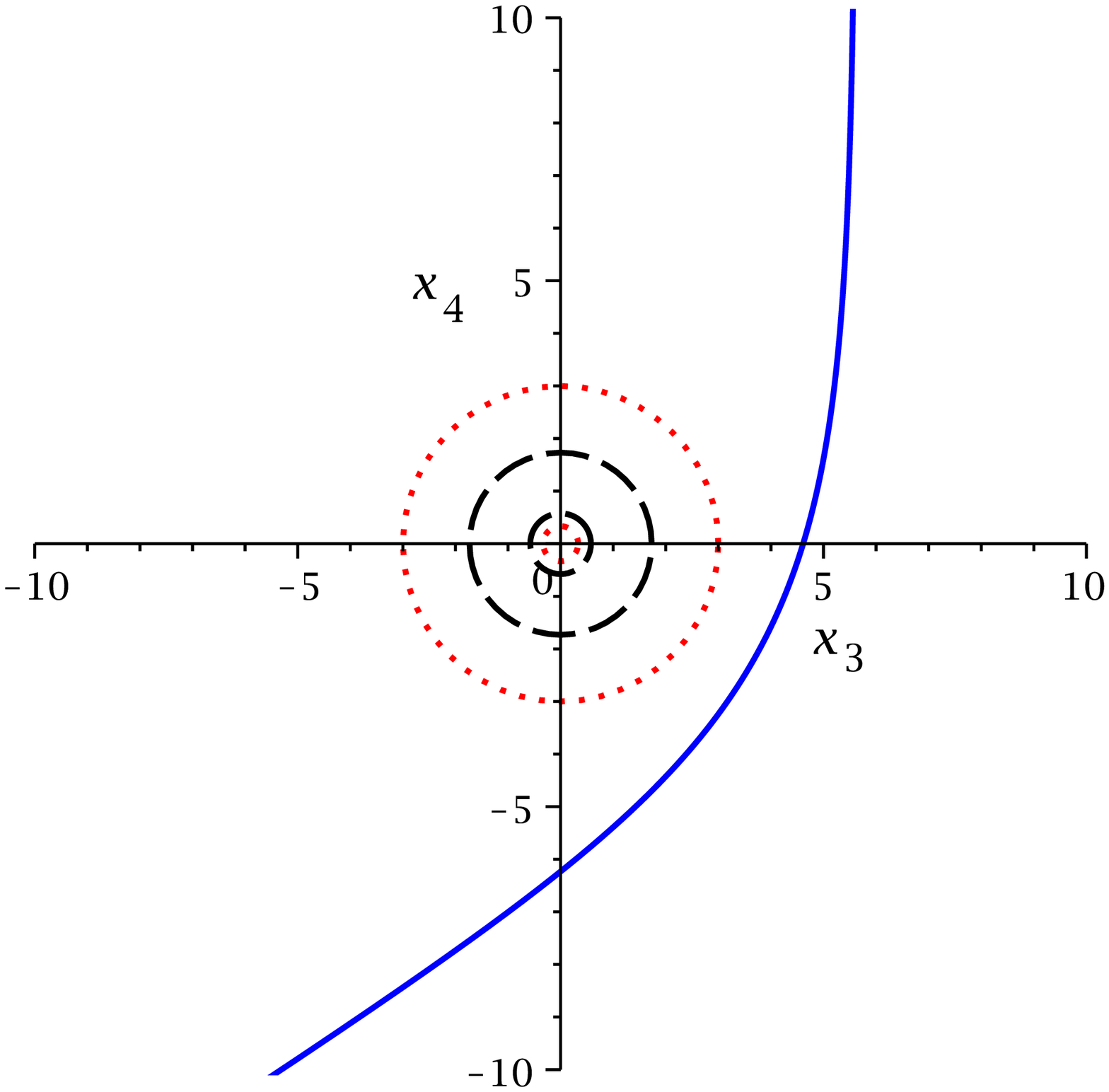}
 }
 \caption{$R=1$, $m=1$, $\lambda=0.5$, $\Psi=5$ and $E=1.6$: Escape orbit on the equatorial plane outside the ring ($x=-1$).}
 \label{pic:sing-phiout-eo}
\end{figure}
\begin{figure}[h]
 \centering
 \subfigure[$a$-$b$-plot\newline 
    The black dashed circles show the position of the horizon and the red dotted circles mark the ergosphere. The orbit is plotted for $\phi=\psi=\frac{\pi}{2}$.]{
   \includegraphics[width=9cm]{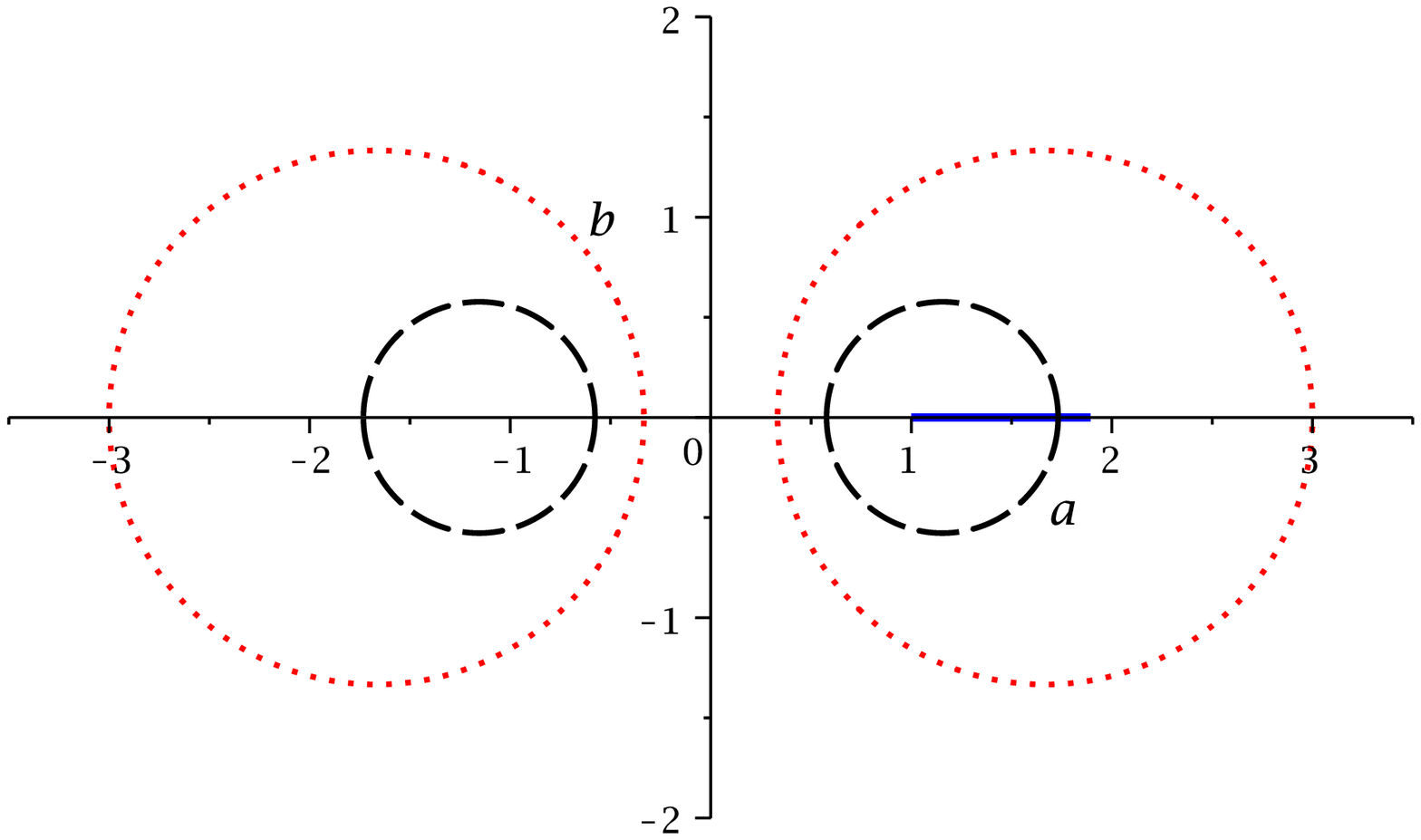}
 }
 \subfigure[Solution $y(\gamma)$\newline 
 The black dashed line shows the position of the event horizon and the black solid line marks the position of the turning point.]{
   \includegraphics[width=6cm]{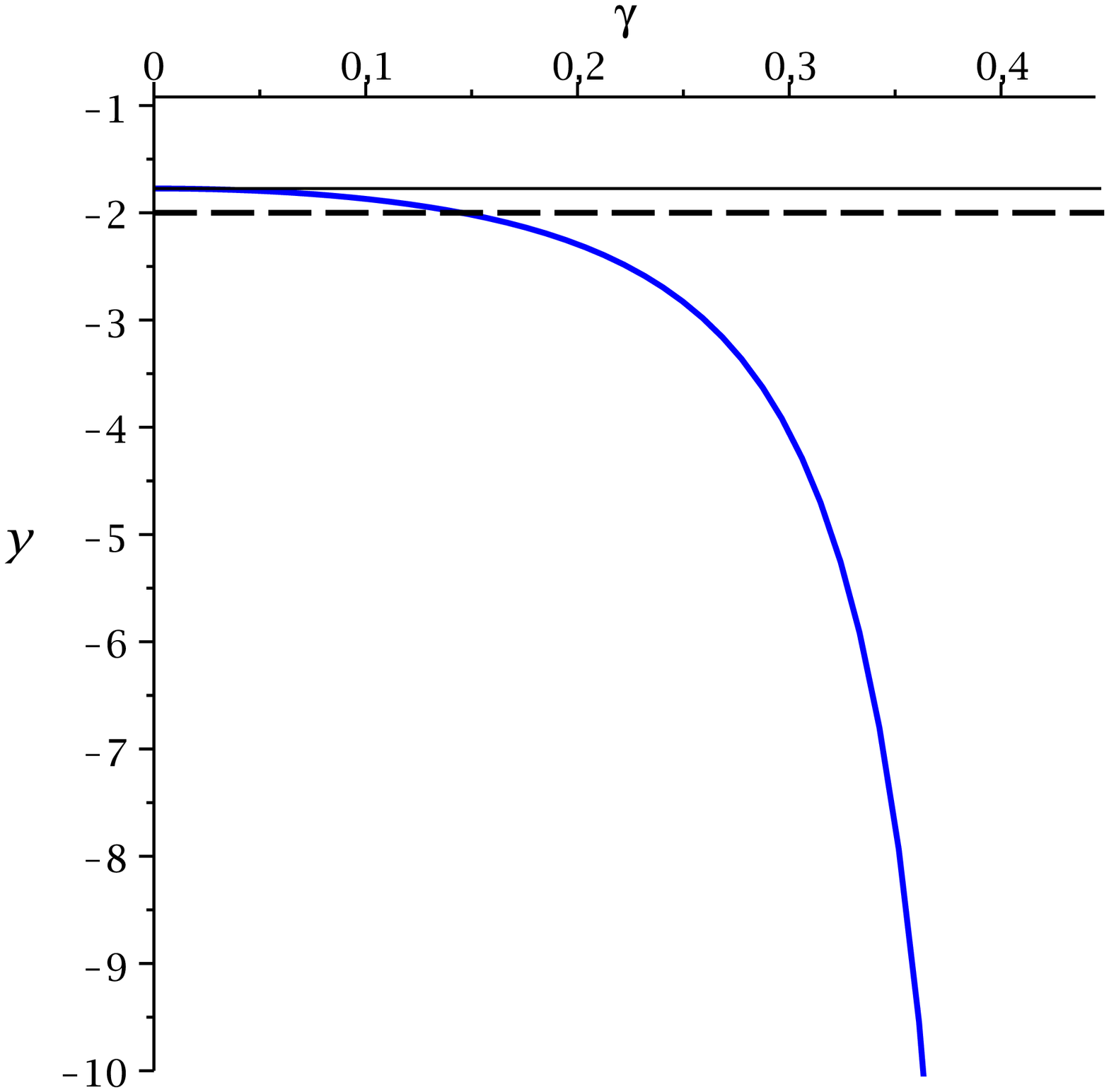}
 }
  \subfigure[ $x_3$-$x_4$-plot ($y$-$\psi$-plane)\newline
    In this plane we are looking at the black ring from above. The black dashed circles show the position of the horizon and the red dotted circles mark the ergosphere. The green solid circle ($\rho_2=1$) is the singularity of the black ring.]{
  \includegraphics[width=6cm]{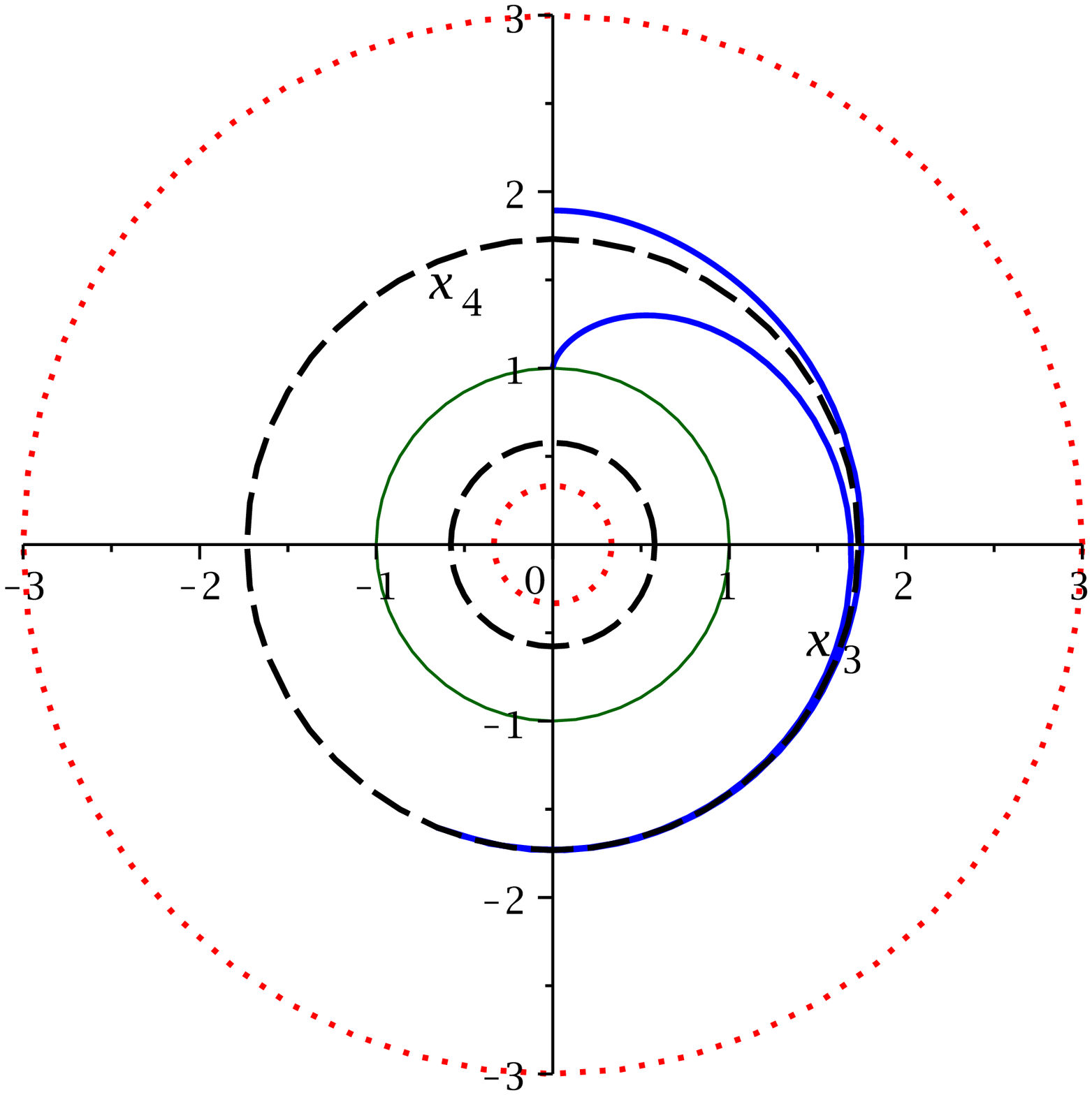}
 }
 \caption{$R=1$, $m=1$, $\lambda=0.5$, $\Psi=5$ and $E=1.6$: Terminating orbit starting at its turning point on the equatorial plane outside the ring ($x=-1$).}
 \label{pic:sing-phiout-to}
\end{figure}

On the equatorial plane around the singly spinning black ring terminating orbits and escape orbits are possible. Figure \ref{pic:sing-phiout-eo} and \ref{pic:sing-phiout-to} show some orbits in $a$-$b$-coordinates (see section \ref{sec:em0-orbit}) and the corresponding solution $y(\gamma)$. Also the $y$-$\psi$-plane is shown in the coordinates $x_3$ and $x_4$ (see section \ref{sec:psi-axis-orbits}).

An escape orbit is depicted in figure \ref{pic:sing-phiout-eo}. Figure \ref{pic:sing-phiout-to} shows a terminating orbit which starts at its turning point and then falls into the singularity.

\begin{figure}[h]
 \centering
 \subfigure[$R=1$, $m=1$, $\lambda=0.5$, $\Psi=5$ and $E=2.75$\newline 
  Here the angular momentum of the particle and the black ring are both positive.]{
   \includegraphics[width=7cm]{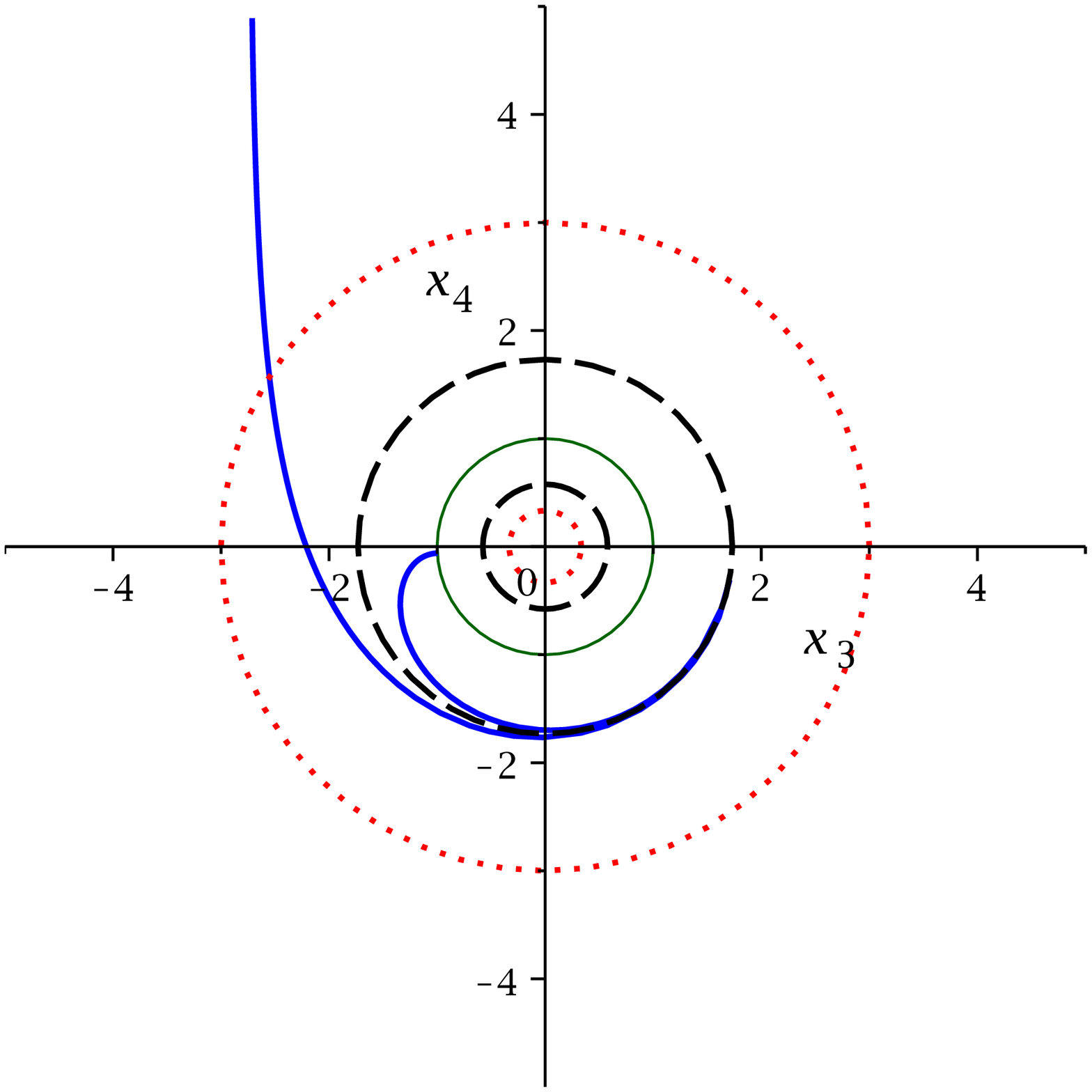}
 }
 \subfigure[$R=1$, $m=1$, $\lambda=0.5$, $\Psi=-5$ and $E=1.6$\newline 
  Here the angular momentum of the particle and the black ring have opposite signs, so the particle changes its direction when approaching the ergosphere.]{
   \includegraphics[width=7cm]{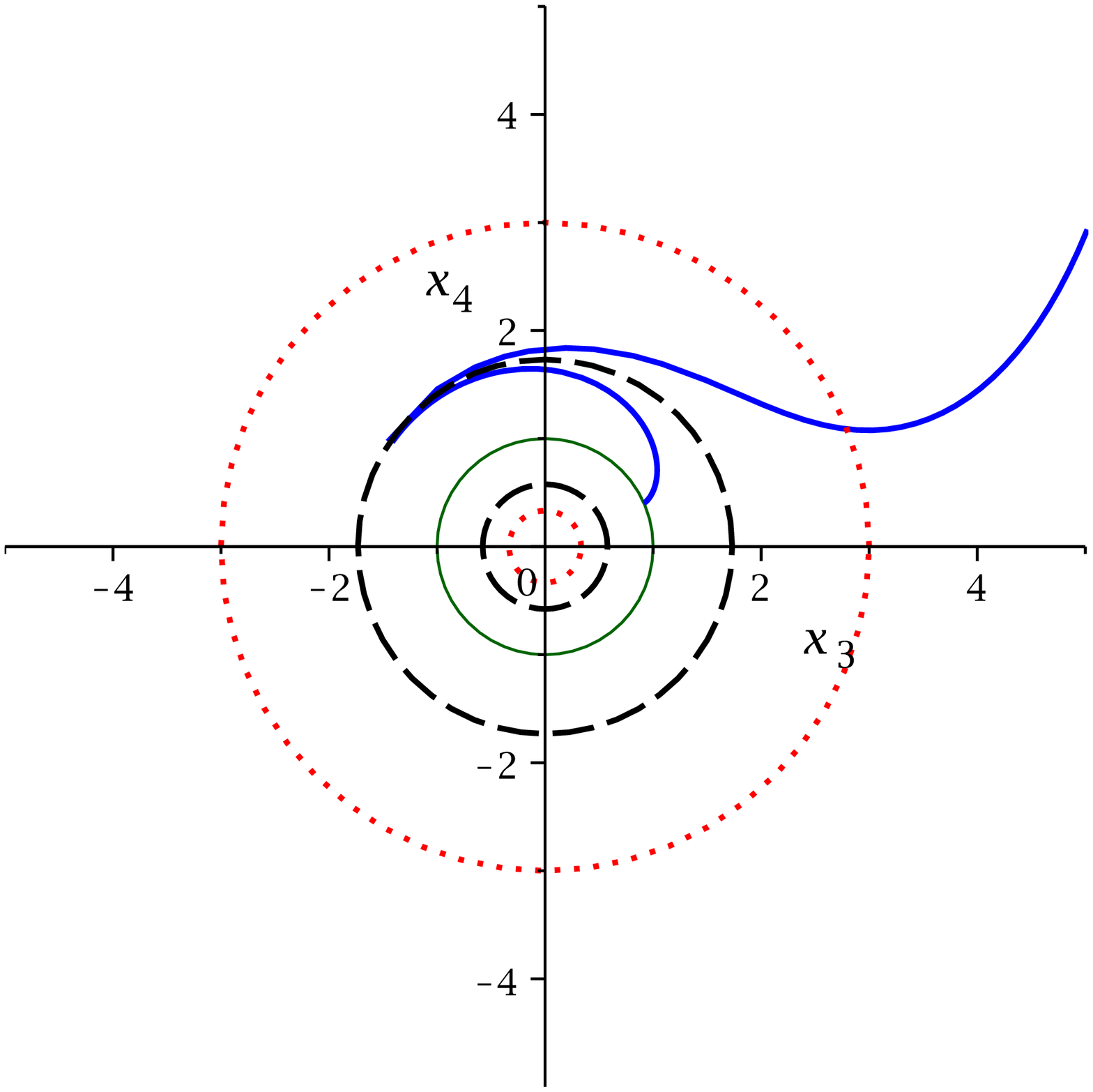}
 }
 \caption{Frame dragging effect: once the particle  enters the ergosphere it is dragged along by the rotation of the black ring.\newline
    The black dashed circles show the position of the horizon and the red dotted circles mark the ergosphere. The green solid circle is the singularity of the black ring.}
 \label{pic:framedragg}
\end{figure}

The frame dragging effect can be seen in figure \ref{pic:framedragg}. Once the particle enters the ergosphere it is dragged along by the rotation of the black ring. If the angular momentum of the particle and the black ring have opposite signs, the particle changes its direction when approaching the ergosphere.
\\

\begin{figure}[h]
 \centering
 \subfigure[$a$-$b$-plot\newline 
    The black dashed circles show the position of the horizon and the red dotted circles mark the ergosphere. The orbit is plotted for $\phi=\psi=\frac{\pi}{2}$.]{
   \includegraphics[width=9cm]{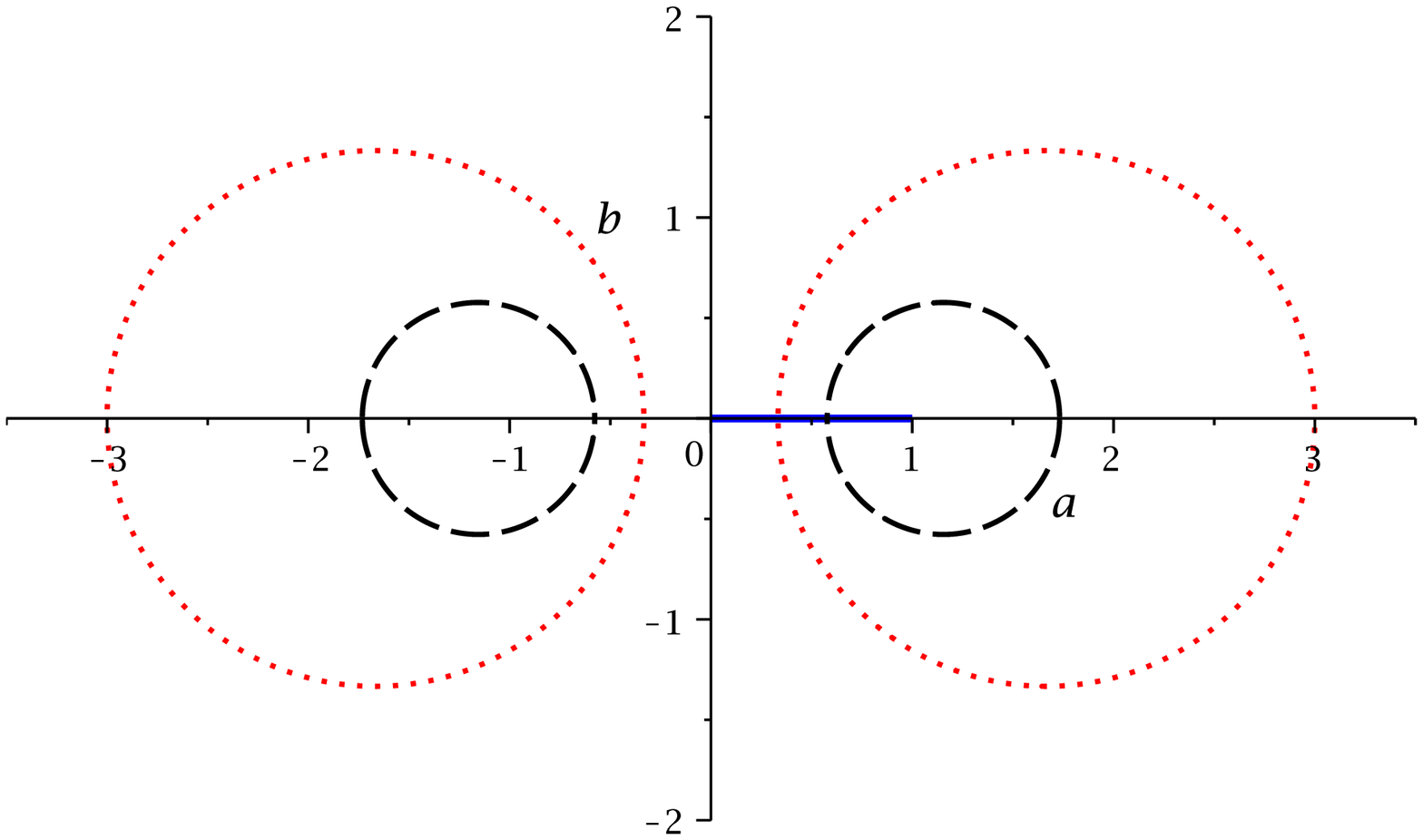}
 }
 \subfigure[Solution $y(\gamma)$\newline 
 The black dashed line shows the position of the event horizon and the black solid line marks the position of the turning point.]{
   \includegraphics[width=6cm]{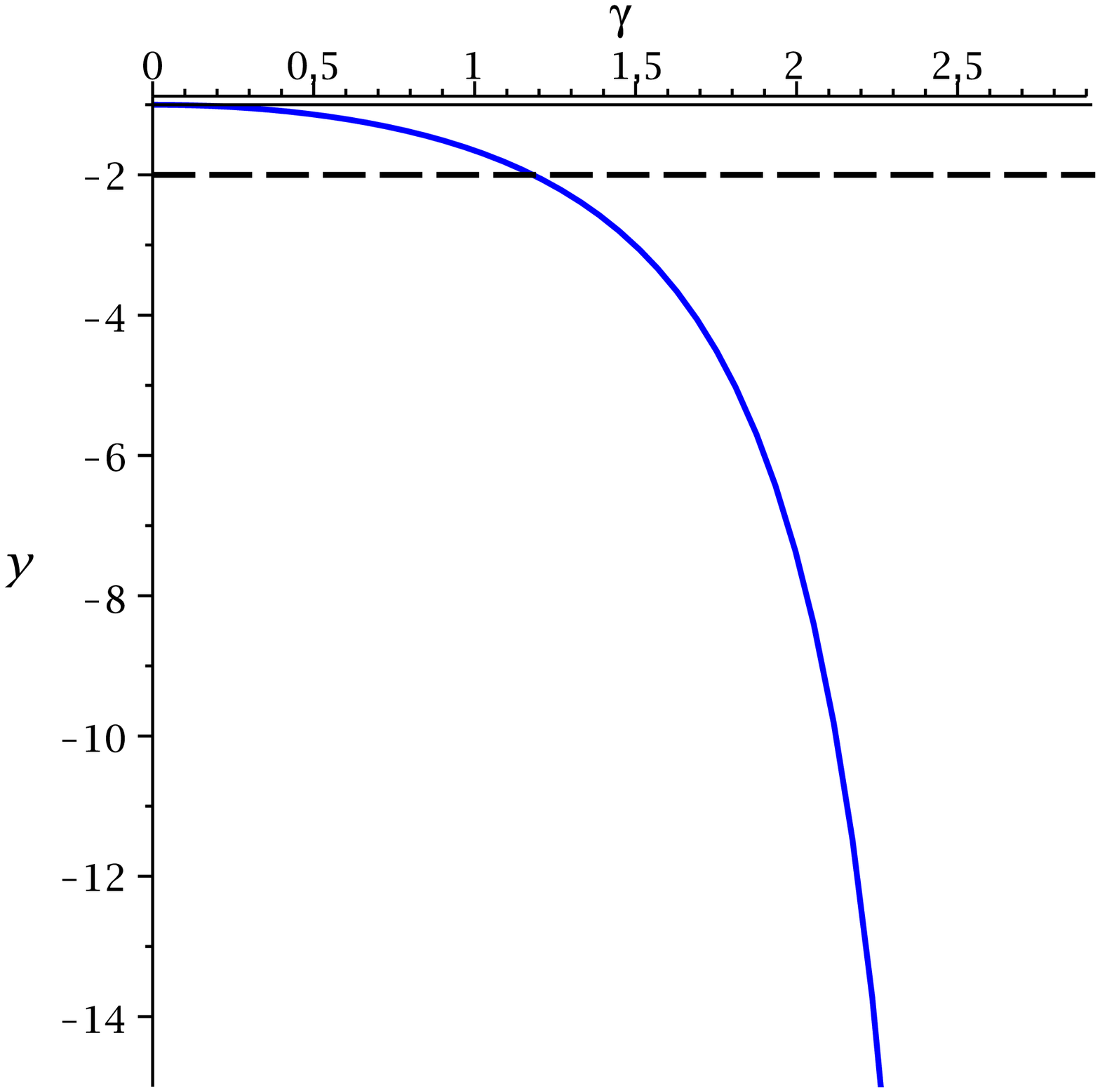}
 }
 \subfigure[$x_3$-$x_4$-plot ($y$-$\psi$-plane)\newline
    In this plane we are looking at the black ring from above. The black dashed circles show the position of the horizon and the red dotted circles mark the ergosphere. The green solid circle ($\rho_2=1$) is the singularity of the black ring.]{
  \includegraphics[width=6cm]{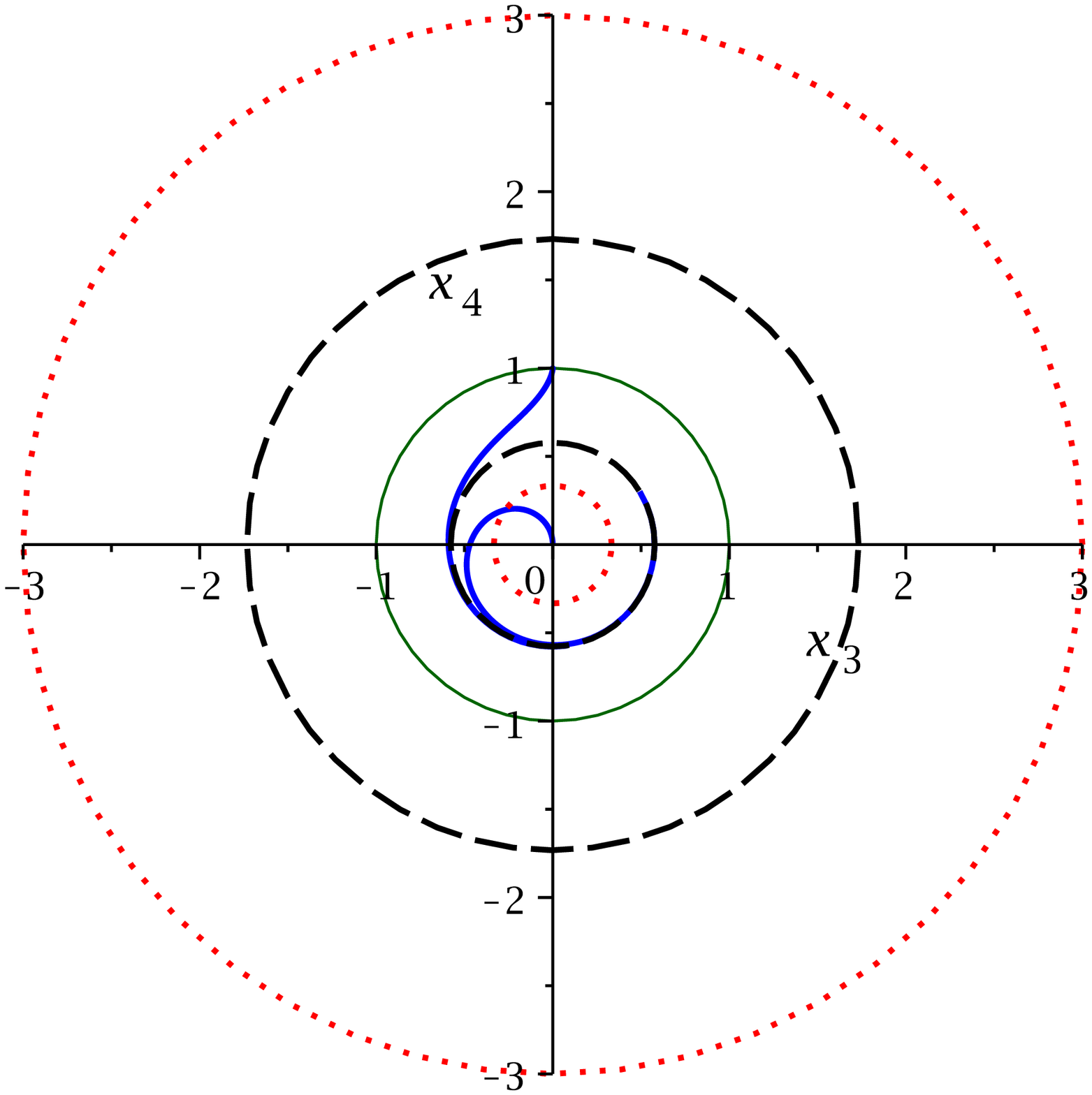}
 }
 \caption{$R=1$, $m=1$, $\lambda=0.5$, $\Psi=0$ and $E=0.8$: Terminating orbit on the equatorial plane inside the ring ($x=+1$). The particle starts at the center of the black ring ($y=-1$,$x=+1$).}
 \label{pic:sing-phiin-to}
\end{figure}

On the equatorial plane enclosed by the black ring only terminating orbits are possible. Figure \ref{pic:sing-phiin-to} shows a terminating orbit which starts at the center of the black ring and then falls into the singularity.

\section{Conclusion}

In this paper we presented the analytical solutions of the geodesic equations of the singly spinning black ring for special cases. Since the Hamilton-Jacobi equation seems not to be separable in general, we had to concentrate on the nullgeodesics in the ergosphere ($E=m=0$), geodesics on the rotational axis ($y=-1$) and geodesics on the equatorial plane ($x=\pm1$).\\
We discussed the general structure of the orbits and gave a complete classification of their types.

In the ergosphere there is just one possible orbit, where light crosses the event horizon and falls inevitably into the singularity (terminating orbit). The $x$-motion bounces back and forth between two values or stays constant at $x=0$, while $y$ ranges from a turning point to $-\infty$.
On the rotational axis $y$ is constant, so here the $x$-motion determines the type of orbit. We found escape orbits and bound orbits, the latter were also shown numerically by Igata et al. \cite{Igata:2010ye}.
On the equatorial plane we found terminating orbits and escape orbits.\\

The separability of the Hamilton-Jacobi equation is a coordinate related phenomenon, so one might think of a coordinate system in which it would be possible to separate the Hamilton-Jacobi equation in general. But recently Igata, Ishihara and Takamori found evidence of chaotic motion in the singly spinning black ring spacetime using the Poincar\'e map \cite{Igata:2010cd}. From that one could conclude that it is not possible to separate the Hamilton-Jacobi equation in any coordinate system.\\

Besides the singly spinning black ring, one could consider black rings with two angular momenta (doubly spinning black ring \cite{Durkee:2008an,Pomeransky:2006bd}) or add charge to the black ring \cite{Elvang:2003yy,Hoskisson:2008qq,Gal'tsov:2009da}. Also a supersymmetric black ring solution was found \cite{Elvang:2004rt}, \cite{Elvang:2004ds}. The methods shown in this paper can be applied to (charged) doubly spinning black rings as well as to supersymmetric black rings. This will be done in future work.

\clearpage

\section{Acknowledgements}

We would like to thank Victor Enolski, Norman G\"urlebeck and Volker Perlick for helpful discussions. We gratefully acknowledge support by the DFG, in particular, also within the DFG Research Training Group 1620 ``Models of Gravity''.


\bibliographystyle{unsrt}

\end{document}